\documentclass[twocolumn,reprint,superscriptaddress,prb,aps,showpacs,msmath,amsmath,amssymb0]{revtex4-2}
\usepackage{graphicx}
\usepackage{dcolumn}
\usepackage{bm}
\usepackage{subcaption}
\usepackage{natbib} 
\usepackage{multirow}
\usepackage{soul}
\usepackage{float}
\usepackage[normalem]{ulem}
\usepackage[usenames,dvipsnames]{color}
\usepackage{appendix}
\usepackage{mathrsfs}
\usepackage[colorlinks=true,linkcolor=blue,citecolor=blue,urlcolor=blue]{hyperref}

\newcommand{\editor}[2]{%
  \expandafter\newcommand\csname #1note\endcsname[1]{%
    \textcolor{#2}{(\textbf{#1:} ##1)}}%
  \expandafter\newcommand\csname #1\endcsname[1]{%
    \textcolor{#2}{##1}}%
  \expandafter\newcommand\csname #1cancel\endcsname[1]{%
    \textcolor{#2}{\sout{##1}}}%
  \expandafter\newcommand\csname #1change\endcsname[2]{%
    \textcolor{#2}{\sout{##1} ##2}}%
  \newenvironment{#1text}{\color{#2}}{\color{black}}
}

\editor{IT}{red}

\begin{document}

\title{Unraveling the electronic structure and the oxygen $K$-edge x-ray absorption near-edge structure spectrum of DyFeO$_3$}

\author{G. Gebreyesus}
\affiliation{Department of Physics, School of Physical and Mathematical Sciences, College of Basic and Applied Sciences, University of Ghana, Ghana}

\author{Eric Macke}
\affiliation{Faculty of Production Engineering, Bremen Center for Computational Materials Science and MAPEX Center for Materials and Processes, Hybrid Materials Interfaces Group, University of Bremen, D-28359 Bremen, Germany}
\affiliation{U Bremen Excellence Chair, University of Bremen, D-28359 Bremen, Germany}

\author{Pietro Delugas}
\affiliation{SISSA, Scuola Internazionale Superiore di Studi Avanzati, via Bonomea 265, 34136 Trieste, Italy}

\author{Weiguo Jing}
\affiliation{UCLouvain, Institute of Condensed Matter and Nanosciences (IMCN),
Chemin des \'Etoiles 8, Louvain-la-Neuve 1348, Belgium}

\author{Banani Biswas}
\affiliation{PSI Center for Neutron and Muon Sciences, Paul Scherrer Institute, 5232 Villigen PSI, Switzerland}

\author{Carlos A. F. Vaz}
\affiliation{Swiss Light Source, Paul Scherrer Institute, 5232 Villigen PSI, Switzerland}

\author{Christof W. Schneider}\email[]{ christof.schneider@psi.ch}
\affiliation{PSI Center for Neutron and Muon Sciences, Paul Scherrer Institute, 5232 Villigen PSI, Switzerland}

\author{Iurii Timrov}\email[]{ iurii.timrov@psi.ch}
\affiliation{PSI Center for Scientific Computing, Theory and Data, Paul Scherrer Institute, 5232 Villigen PSI, Switzerland}

\begin{abstract} 
Rare-earth orthoferrites such as DyFeO$_3$ exhibit a rich interplay between localized rare-earth and transition-metal moments, giving rise to complex magnetic phases and magnetoelectric phenomena. Understanding their electronic and spectroscopic properties from first principles requires an accurate description of the localized Fe-$3d$ and Dy-$4f$ states and their hybridization with O-$2p$ states. Here, we combine first-principles calculations and x-ray absorption near-edge structure (XANES) measurements to investigate the electronic structure and the O $K$-edge spectrum of DyFeO$_3$. We employ density-functional theory (DFT) with Hubbard $U$ corrections (DFT+$U$) determined from first principles using density-functional perturbation theory, the HSE06 hybrid functional, and orbital-resolved DFT+$U$ with Hubbard parameters calibrated to reproduce the electronic structure computed using HSE06. We find that standard DFT+$U$, despite improving the band gap, substantially underestimates the crystal-field splitting of the unoccupied Fe-$3d$ states and consequently fails to accurately reproduce the separation of the two lowest-energy features in the O $K$-edge spectrum. HSE06 provides a more balanced description of the relevant electronic states, including the Fe-$3d$ crystal-field splitting. Mapping the corresponding electronic structure onto orbital-resolved DFT+$U$ yields a spectrum that remarkably reproduces the two lowest-energy experimental features and captures the main characteristics of the spectrum at higher energies. These results establish the low-energy O $K$-edge features as a sensitive probe of the Fe-$3d$ crystal-field splitting, while showing that the localized Dy-$4f$ states leave no distinct spectral fingerprints despite their importance for the magnetic properties. More broadly, this work demonstrates that orbital-resolved DFT+$U$ can provide an efficient route to simulate reliable XANES spectra and relate its spectral features to the electronic structure of the material.
\end{abstract}

\date{\today} 

\maketitle

\section{Introduction}
\label{sec:intro}

Perovskite-type oxides $RMO_3$, where $R$ denotes a rare-earth or alkaline-earth cation and $M$ a transition-metal cation, constitute a broad and extensively studied class of inorganic materials that exhibit remarkable structural flexibility, diverse magnetic ground states, and a wide range of electronic and functional properties~\cite{Walton:2020}. Among rare-earth orthoferrites, DyFeO$_3$ is of particular interest because of the strong interplay between its Fe$^{3+}$ and Dy$^{3+}$ magnetic sublattices, which gives rise to a rich sequence of magnetic phases, spin-reorientation phenomena, and pronounced magnetoelectric effects~\cite{Tokunaga:2008, Afanasiev:2021, Li:2025}. In particular, magnetic-field-induced ferroelectricity has been demonstrated at low temperatures, with the electric polarization arising from the coupling between the Fe and Dy magnetic sublattices through exchange striction~\cite{Tokunaga:2008, Rajeswaran:2013}. These properties make DyFeO$_3$ an attractive platform for studying the interplay between magnetic anisotropy, spin-orbit coupling, and magnetoelectric phenomena.

\begin{figure}[t]
    \centering
    \includegraphics[width=0.98\linewidth]{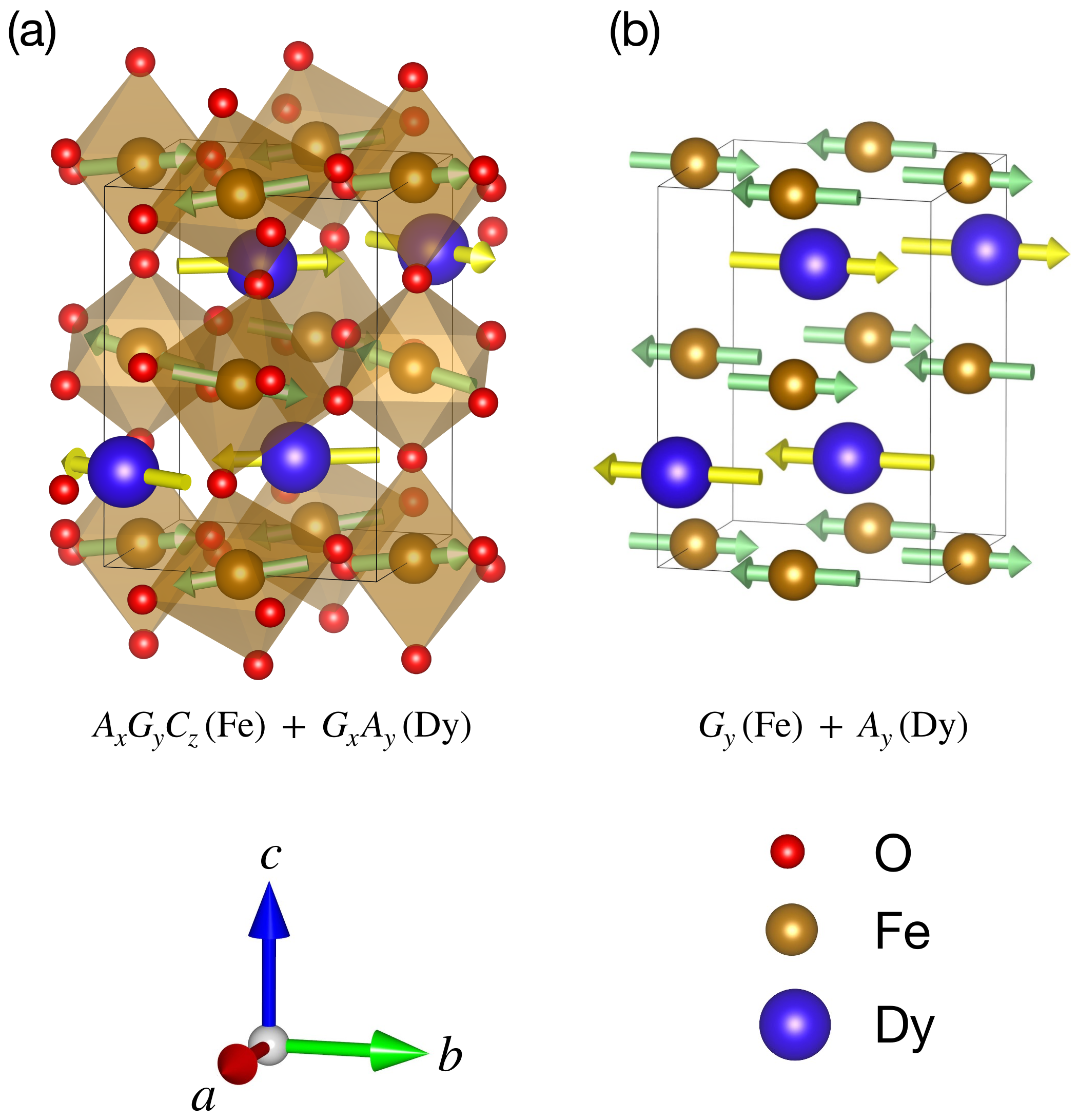}
    \caption{Crystal structure and magnetic configurations of DyFeO$_3$. The orthorhombic unit cell contains four formula units and belongs to the space group $Pbnm$. (a)~Noncollinear commensurate magnetic configuration, consisting of an $A_xG_yC_z$ magnetic structure on the Fe sublattice and a $G_xA_y$ structure on the Dy sublattice. The Dy magnetic moments are tilted by $33^\circ$ from the crystallographic $b$ axis. The $A_x$ and $C_z$ components for the Fe sublattice are intentionally exaggerated for visualization purposes and do not represent experimentally determined moment amplitudes. The magnetic structures are described using Bertaut notation~\cite{Bertaut:1963, Bertaut_explanation}. (b)~Corresponding simplified collinear magnetic configuration, in which all Fe and Dy moments lie along the $b$ axis. The Fe sublattice exhibits $G$-type antiferromagnetic ordering ($G_y$), while the Dy sublattice exhibits $A$-type antiferromagnetic ordering ($A_y$). Thus, the $A_x$ and $C_z$ components for Fe and the $G_x$ component for Dy, that are present in the noncollinear configuration in (a), are omitted. For clarity, the FeO$_6$ octahedra and O atoms are omitted in (b). The crystallographic axes and the atomic color code are indicated at the bottom of the figure. Structures were rendered using VESTA~\cite{Momma:2011}.}
    \label{fig:Crystal_structure}
\end{figure}

DyFeO$_3$ crystallizes in the centrosymmetric orthorhombic structure ($Pbnm$ space group) and exhibits magnetic ordering over a broad temperature range. The Fe$^{3+}$ moments order antiferromagnetically below $T_\mathrm{N}^{\mathrm{Fe}}\sim645$~K, with the dominant $G$-type component oriented along the [100] direction ($a$ axis)~\cite{Biswas:2022}. Spin-orbit coupling induces a small canting of the Fe moments, giving rise to a weak ferromagnetic component along the [001] direction ($c$ axis). This magnetic structure is denoted as $G_xA_yF_z$ in Bertaut's notation~\cite{Bertaut:1963, Bertaut_explanation}. Upon cooling through $T_\mathrm{SR}\sim52$~K, DyFeO$_3$ undergoes a Morin-type spin-reorientation transition, during which the Fe magnetic structure changes from $G_xA_yF_z$ to $A_xG_yC_z$~\cite{Gorodetsky:1968, Li:2021, Biswas:2022}. The dominant antiferromagnetic component therefore rotates from the $a$ axis to the $b$ axis, while the weak ferromagnetic component along the $c$ axis disappears. This spin-reorientation transition reflects the delicate balance between competing magnetic anisotropies and the interactions between the Fe$^{3+}$ and Dy$^{3+}$ sublattices~\cite{Cui:2024}. At lower temperatures, the Dy$^{3+}$ moments develop long-range magnetic correlations below $T_\mathrm{N}^{\mathrm{Dy}}\approx4.2$~K~\cite{Zhao:2014}. Earlier studies described this order in terms of a noncollinear $G_xA_y$ structure, with the Dy moments predominantly confined to the $ab$ plane and their easy axes tilted by approximately $33^\circ$ from the $b$ axis~\cite{Cui:2024}. More recent high-resolution neutron-diffraction measurements, however, revealed that the low-temperature Dy order in zero magnetic field is incommensurate, while the Fe sublattice retains its commensurate $A_xG_yC_z$ structure~\cite{Biswas:2022, Ritter:2022}. This complex evolution of the Fe and Dy magnetic order, together with their strong mutual coupling, underlies the unusual low-temperature magnetoelectric response of DyFeO$_3$ and makes the material a particularly stringent test case for first-principles descriptions of its electronic, magnetic, and spectroscopic properties.

A first-principles description of the electronic structure of DyFeO$_3$ within density-functional theory (DFT)~\cite{Hohenberg:1964,Kohn:1965} is challenging because of the localized Fe-$3d$ and Dy-$4f$ electrons and the complex interplay between the two magnetic sublattices. Standard exchange-correlation (xc) functionals, such as the local-density approximation (LDA)~\cite{Kohn:1965} and generalized-gradient approximation (GGA)~\cite{Perdew:1992}, suffer from self-interaction errors~\cite{Perdew:1981,MoriSanchez:2006}, which can lead to an inadequate description of localized $d$ and $f$ states and, consequently, of their energies, hybridization, and magnetic properties. Furthermore, previous studies have demonstrated that treating the Dy-$4f$ states as core states can substantially affect the predicted magnetic properties of DyFeO$_3$, highlighting the importance of including these states explicitly in the valence manifold of the pseudopotential-based DFT calculations~\cite{Stroppa:2010, Ameri:2021, Cui:2024}. A widely used approach to improve the description of localized electronic states is DFT augmented with an on-site Hubbard correction (DFT+$U$)~\cite{Anisimov:1991, Dudarev:1998, Himmetoglu:2014, Kulik:2006, Kulik:2008}. A central challenge of DFT+$U$, however, is the determination of the Hubbard parameters for the electronic states to which the correction is applied. Previous DFT+$U$ studies of DyFeO$_3$ have applied Hubbard corrections to both the Fe-$3d$ and Dy-$4f$ states, with the corresponding $U$ parameters determined empirically by fitting selected experimental observables, such as magnetic moments or structural parameters~\cite{Stroppa:2010,Ameri:2021,Cui:2024}; the reported values are summarized in Table~\ref{tab:Hub_param}. More computationally demanding approaches, including hybrid functionals such as the Heyd-Scuseria-Ernzerhof (HSE) functional~\cite{Heyd:2003, Heyd:2006} and the $GW$ method~\cite{Hedin:1965}, have also been applied to DyFeO$_3$~\cite{Stroppa:2010}. However, our survey of the literature shows that these approaches have so far been used only for analyzing the density of occupied states, without a systematic analysis of the unoccupied states. Therefore, a comprehensive assessment of the electronic structure of DyFeO$_3$ based on DFT+$U$ with \textit{ab initio} Hubbard parameters, together with a systematic comparison to HSE, is still lacking. This is essential for understanding the relative energies, orbital character, and hybridization of the Fe-$3d$, Dy-$4f$, and O-$2p$ states. The reliability of a calculated electronic structure can be assessed most directly by comparison with experimental spectroscopic measurements. While photoemission spectroscopy provides access to the occupied electronic states, the unoccupied states can be probed by X-ray absorption spectroscopy. In particular, the O $K$-edge X-ray absorption near-edge structure (XANES) spectrum probes transitions from the O-$1s$ core level into unoccupied states with O-$2p$ character and is therefore sensitive to the hybridization of O-$2p$ states with the Fe-$3d$ and Dy-$4f$ states~\cite{Frati:2020}. Experimental O $K$-edge XANES spectra of DyFeO$_3$ have been reported~\cite{Chiang:2011} and are reexamined here, providing an important experimental benchmark for the calculated unoccupied electronic structure. However, to the best of our knowledge, a first-principles investigation of the O $K$-edge XANES spectrum of DyFeO$_3$ is still lacking.

Here, we present a comprehensive first-principles investigation of the electronic structure and O $K$-edge XANES spectrum of DyFeO$_3$, combining DFT+$U$, HSE06, orbital-resolved DFT+$U$ (OR-DFT+$U$), and experiment. We explicitly include the Dy-$4f$ states in the valence manifold and determine the Fe-$3d$ and Dy-$4f$ Hubbard parameters of DFT+$U$ self-consistently using linear-response theory~\cite{Cococcioni:2005} within density-functional perturbation theory (DFPT)~\cite{Timrov:2018, Timrov:2021}. We find that, although this approach substantially improves the band gap, the resulting electronic structure does not accurately reproduce the Fe-$3d$ crystal-field splitting relevant to the O $K$-edge spectrum. We therefore use HSE calculations with varying fractions of exact exchange $\alpha$ and find that the conventional HSE06 functional ($\alpha=0.25$) provides a good description of the relevant electronic states. Since hybrid functionals are not currently supported in our XANES calculations, we use HSE06 as a reference to calibrate OR-DFT+$U$, in which separate Hubbard parameters are assigned to individual Fe-$3d$ and Dy-$4f$ submanifolds. The resulting electronic structure closely reproduces the main features of the HSE06 PDOS and yields an O $K$-edge XANES spectrum that accurately reproduces the two lowest-energy experimental features associated primarily with O-$2p$ and Fe-$3d$ hybridization. The remaining discrepancies at higher energies point to limitations in the description of the Dy-$5d$ and other delocalized unoccupied states. Our results demonstrate that OR-DFT+$U$ provides an efficient route to XANES simulations when hybrid-functional-based XANES calculations are not directly available.

The remainder of the paper is organized as follows. Section~\ref{sec:methods} summarizes the computational methods employed in this work, including DFT+$U$, the DFPT approach for determining Hubbard parameters, orbital-resolved DFT+$U$, the HSE hybrid functional, and the methodology used to calculate XANES spectra. This section also describes the computational and experimental details. Section~\ref{sec:Results_and_Discussion} presents and discusses the results, starting with the electronic structure and followed by the O $K$-edge XANES spectra and their comparison with experiment. Finally, Sec.~\ref{sec:conclusions} summarizes the findings of this work.

\section{Methods}
\label{sec:methods}

\subsection{Computational approach}
\label{sec:comput_approach}

In this section, we briefly summarize the computational approaches employed to calculate the electronic structure and XANES spectra of DyFeO$_3$. For simplicity, the formalism is presented within the collinear spin-polarized framework of norm-conserving pseudopotentials.

\subsubsection{Calculation of XANES spectra}

Within Fermi's golden rule, the X-ray absorption cross section can be written as
\begin{equation}
    \sigma(\omega) =
    4\pi^2\alpha_0\hbar\omega
    \sum_{\mathrm{f},\mathbf{k},\sigma}
    \left|
    M^\sigma_{\mathrm{i} \rightarrow \mathrm{f},\mathbf{k}}
    \right|^2 
    \delta\left(
    \varepsilon^\sigma_{\mathrm{f},\mathbf{k}}
    -
    \varepsilon^\sigma_{\mathrm{i},\mathbf{k}}
    -
    \hbar\omega
    \right),
    \label{eq:cross_section}
\end{equation}
where $\alpha_0$ is the fine-structure constant, $\hbar\omega$ is the photon energy, $\mathbf{k}$ denotes a point in the Brillouin zone, and $\sigma$ is the spin index. The transition amplitudes are given by the matrix elements
\begin{equation}
    M^\sigma_{\mathrm{i} \rightarrow \mathrm{f},\mathbf{k}}
    =
    \left\langle
    \psi^\sigma_{\mathrm{f},\mathbf{k}}
    \left|
    \hat{D}
    \right|
    \psi^\sigma_{\mathrm{i},\mathbf{k}}
    \right\rangle ,
\end{equation}
where $\hat{D}$ is the transition operator. Within the electric-dipole
approximation,
\begin{equation}
    \hat{D} = \boldsymbol{\epsilon} \cdot \hat{\mathbf{r}},
\end{equation}
where $\boldsymbol{\epsilon}$ is the unit polarization vector of the
incident photon and $\hat{\mathbf{r}}$ is the electron position operator.
The quantities $\psi^\sigma_{\mathrm{i},\mathbf{k}}$ and $\psi^\sigma_{\mathrm{f},\mathbf{k}}$ are the initial and final electronic states, with corresponding energies $\varepsilon^\sigma_{\mathrm{i},\mathbf{k}}$ and $\varepsilon^\sigma_{\mathrm{f},\mathbf{k}}$.
The cross section in Eq.~\eqref{eq:cross_section} can be evaluated efficiently using a recursive Lanczos algorithm, which recasts the spectral function as a continued fraction and thereby avoids the explicit calculation of a large number of unoccupied electronic states~\cite{Taillefumier:2002,Gougoussis:2009b}.

The electronic states entering Eq.~\eqref{eq:cross_section} are obtained by solving the spin-polarized Kohn-Sham equations,
\begin{equation}
    \hat{H}^{\sigma}
    \psi^\sigma_{v,\mathbf{k}}
    =
    \varepsilon^\sigma_{v,\mathbf{k}}
    \psi^\sigma_{v,\mathbf{k}},
    \label{eq:KS}
\end{equation}
where $v$ is the electronic band index and $\hat{H}^{\sigma}$ is the
spin-dependent Kohn-Sham Hamiltonian of the ground-state electronic system. As was stated earlier, for transition-metal oxides containing localized $d$ and $f$ electrons, standard LDA or GGA functionals can suffer from self-interaction errors, leading to an excessive delocalization of these states and an inaccurate description of their energies and hybridization. Since the XANES spectrum is directly sensitive to the unoccupied electronic structure, such inaccuracies can propagate into the calculated absorption spectrum. To improve the description of localized states, we therefore employ DFT+$U$~\cite{Anisimov:1991, Dudarev:1998, Himmetoglu:2014} and the HSE hybrid functional~\cite{Heyd:2003, Heyd:2006} as described below.

\subsubsection{Standard DFT+$U$}
\label{sec:standard_dftu}

In the ``standard'' DFT+$U$ approach, the spin-dependent Kohn-Sham
Hamiltonian is written as
\begin{equation}
    \hat{H}^\sigma =
    \hat{H}^\sigma_{\mathrm{DFT}} +
    \hat{V}^\sigma_U ,
    \label{eq:Hamiltonian_global_U}
\end{equation}
where $\hat{H}^\sigma_{\mathrm{DFT}}$ is the DFT Hamiltonian and
$\hat{V}^\sigma_U$ is the Hubbard potential associated with the on-site
Hubbard correction. Within the simplified rotationally-invariant formulation of Dudarev~\cite{Dudarev:1998}, this potential is given by
\begin{equation}
    \hat{V}^\sigma_U =
    \sum_I \sum_{m m'}
    U^{I}
    \left(
    \frac{\delta_{m m'}}{2}
    -
    n^{I\sigma}_{m m'}
    \right)
    |\varphi^I_m\rangle
    \langle\varphi^I_{m'}| ,
    \label{eq:potU}
\end{equation}
where $I$ labels the atomic sites to which the Hubbard correction is applied, while $m$ and $m'$ label the magnetic quantum numbers of the selected shell with the orbital quantum number $l$. In the present case, $l=2$ for the Fe-$3d$ shell and $l=3$ for the Dy-$4f$ shell. The parameter $U^I$ is the effective on-site Hubbard parameter and, within standard DFT+$U$, it is a single parameter shared by all orbitals belonging to the selected shell of atom $I$.

The quantities $\varphi^I_m$ are localized atom-centered orbitals, and
$n^{I\sigma}_{m m'}$ is the corresponding occupation matrix,
\begin{equation}
   n^{I\sigma}_{m m'} =
   \sum_{v,\mathbf{k}}
   f^\sigma_{v,\mathbf{k}}
   \langle \psi^\sigma_{v,\mathbf{k}} | \varphi^I_{m'} \rangle
   \langle \varphi^I_m | \psi^\sigma_{v,\mathbf{k}}\rangle ,
   \label{eq:occ_matrix}
\end{equation}
where $f^\sigma_{v,\mathbf{k}}$ denotes the occupation of the Kohn-Sham
state $\psi^\sigma_{v,\mathbf{k}}$. By diagonalizing the occupation matrix,
\begin{equation}
    \sum_{m'}
    n^{I\sigma}_{m m'}
    \nu^{I\sigma}_{m'i}
    =
    \lambda^{I\sigma}_i
    \nu^{I\sigma}_{mi},
    \label{eq:occ_matrix_eigen}
\end{equation}
one obtains its eigenvalues $\lambda^{I\sigma}_i$ and eigenvectors
$\nu^{I\sigma}_{mi}$. The corresponding rotated orbitals (eigen-orbitals) are defined as
\begin{equation}
    |\phi^{I\sigma}_i\rangle =
    \sum_m \nu^{I\sigma}_{mi} \, |\varphi^I_m\rangle .
    \label{eq:rotated_orbitals}
\end{equation}
In this basis, the Hubbard potential can be expressed in diagonal form as
\begin{equation}
    \hat{V}^\sigma_U =
    \sum_I U^I \sum_i
    \left(
    \frac{1}{2} - \lambda^{I\sigma}_i
    \right)
    |\phi^{I\sigma}_i\rangle
    \langle\phi^{I\sigma}_i| .
    \label{eq:potU_OR_2}
\end{equation}
This representation makes explicit that, although the Hubbard potential acts differently on individual orbitals through their occupation eigenvalues $\lambda^{I\sigma}_i$, the same Hubbard parameter $U^I$ is assigned to all orbitals within a given shell.

\subsubsection{Orbital-resolved DFT+$U$}
\label{sec:or_dftu}

The OR-DFT+$U$ approach~\cite{Macke:2024} extends this formulation by allowing the Hubbard parameter to depend on the individual eigen-orbitals of the occupation matrix, $\phi^{I\sigma}_i(\mathbf{r})$ [see Eq.~\eqref{eq:rotated_orbitals}]. The corresponding Hamiltonian is
\begin{equation}
    \hat{H}^\sigma =
    \hat{H}^\sigma_{\mathrm{DFT}} +
    \hat{V}^\sigma_{U,\mathrm{OR}},
    \label{eq:Hamiltonian_OR}
\end{equation}
with
\begin{equation}
    \hat{V}^\sigma_{U,\mathrm{OR}} =
    \sum_I \sum_i
    U^I_i
    \left(
    \frac{1}{2} - \lambda^{I\sigma}_i
    \right)
    |\phi^{I\sigma}_i\rangle
    \langle\phi^{I\sigma}_i| .
    \label{eq:potU_OR_3}
\end{equation}
The essential difference between Eqs.~\eqref{eq:potU_OR_2} and
\eqref{eq:potU_OR_3} is that standard DFT+$U$ employs a single parameter
$U^I$ for the entire shell of atom $I$, whereas OR-DFT+$U$
assigns an independent parameter $U^I_i$ to each eigen-orbital
$|\phi^{I\sigma}_i\rangle$. This additional flexibility allows the Hubbard correction to account for orbital-dependent variations in the electronic structure that cannot, in general, be represented by a single shell-averaged Hubbard parameter.

In the present work, we combine the OR-DFT+$U$ formalism with the calculation of XANES spectra in Eq.~\eqref{eq:cross_section}. This implementation is available in the public version of \textsc{Quantum ESPRESSO} starting from version 7.5. The O $K$-edge XANES spectra reported in this work are calculated using this new implementation.

\subsubsection{Determination of $U$}

The Hubbard parameters entering DFT+$U$ are not known {\it a priori} and are therefore often determined semiempirically. One common approach is to adjust $U$ to reproduce selected experimental properties (e.g. lattice parameters, magnetic moments, etc.)~\cite{Stroppa:2010, Ameri:2021, Cui:2024}. Alternatively, $U$ can be tuned such that DFT+$U$ reproduces the results of more computationally demanding methods, such as hybrid-functional calculations using the HSE functional~\cite{Yu:2020}. These procedures are inherently nonunique, as different target properties or reference methods may yield different optimal values of $U$, resulting in a spread of DFT+$U$ predictions. In the present work, we employ two different strategies depending on the DFT+$U$ formulation. For standard DFT+$U$, the Hubbard parameters are determined from first principles using DFPT as described below. For OR-DFT+$U$, we instead determine the orbital-resolved Hubbard parameters by tuning them to reproduce the positions of the Fe-$3d$ and Dy-$4f$ states obtained from HSE06 calculations, as reflected in the corresponding PDOS.

The first-principles determination of $U$ can be performed using several approaches, including constrained random-phase approximation (cRPA)~\cite{Springer:1998, Kotani:2000, Aryasetiawan:2004}, Hartree-Fock-based ACBN0~\cite{Agapito:2015, Lee:2020, TancogneDejean:2020}, and linear-response theory~\cite{Cococcioni:2005}. Here, we determine $U$ using the method of Ref.~\cite{Cococcioni:2005}, which is based on enforcing the piecewise linearity of the total energy with respect to the number of electrons in the Hubbard manifold. Specifically, $U$ is given by the diagonal elements of an effective interaction matrix defined as the difference between the inverse bare and screened response matrices:
\begin{equation}
    U^I = \left(\chi_0^{-1} - \chi^{-1}\right)_{II} \,,
    \label{eq:Ucalc}
\end{equation}
where $\chi_0$ and $\chi$ are response matrices describing the variation of the atomic occupations in response to potential shifts applied to individual Hubbard manifolds. More specifically,
\begin{equation}
    \chi_{IJ}
    =
    \sum_{m\sigma}
    \frac{d n^{I\sigma}_{mm}}{d\alpha^J} \,,
\end{equation}
where $\alpha^J$ is the amplitude of the perturbing potential applied to the Hubbard manifold associated with site $J$. The screened response $\chi$ is evaluated after self-consistent solution of the Kohn-Sham equations, whereas the bare response $\chi_0$ is evaluated before the self-consistent rearrangement of the Hartree and exchange-correlation potentials~\cite{Cococcioni:2005}. Their difference therefore accounts for the screening of the applied perturbation by the electronic system.

A direct linear-response calculation requires applying localized perturbations to individual atoms in a supercell. The DFPT formulation avoids this computationally demanding step: the response to a localized perturbation in a supercell is reconstructed as a sum over the responses to monochromatic perturbations in the primitive cell, with wave vectors $\mathbf{q}$ sampled on a regular grid of $N_{\mathbf q}$ points in the Brillouin zone~\cite{Timrov:2018}:
\begin{equation}
    \frac{d n^{I\sigma}_{mm'}}{d\alpha^J}
    =
    \frac{1}{N_{\mathbf q}}
    \sum_{\mathbf q}^{N_{\mathbf q}}
    e^{i\mathbf q\cdot(\mathbf R_c-\mathbf R_{c'})}
    \Delta_{\mathbf q}^{s'} n^{s\sigma}_{mm'} \,.
    \label{eq:dnq}
\end{equation}
Here, the atomic indices $I$ and $J$ are resolved into atomic ($s$ and $s'$) and unit-cell ($c$ and $c'$) labels, such that $I\equiv(c,s)$ and $J\equiv(c',s')$, while $\mathbf R_c$ and $\mathbf R_{c'}$ are the corresponding Bravais lattice vectors. The quantities $\Delta_{\mathbf q}^{s'} n^{s\sigma}_{mm'}$ are the lattice-periodic responses of the atomic occupations to a monochromatic perturbation with wave vector $\mathbf q$. They are obtained by solving the DFPT equations independently for each $\mathbf q$ point~\cite{Timrov:2018}. The $\mathbf q$-point grid is chosen sufficiently dense so that the reconstructed localized perturbation is effectively decoupled from its periodic replicas. Thus, the supercells required in conventional linear-response calculations are replaced by a set of independent monochromatic perturbations calculated in the primitive cell. This formulation has enabled efficient first-principles calculations of a broad range of materials' properties, as demonstrated in several recent studies~\cite{Mahajan:2022, Gebreyesus:2023, Timrov:2023, Binci:2023, Haddadi:2024, Gelin:2024, Bonfa:2024, Chang:2025, dosSantos:2025, Binci:2025}. 
The linear-response formulation for determining the shell-averaged Hubbard parameter $U^I$ has also been generalized to obtain orbital-resolved Hubbard parameters $U_i^I$. The details of this formulation are given in Ref.~\cite{Macke:2024}.

\subsubsection{HSE hybrid functional}

For the HSE hybrid functional, the Kohn-Sham Hamiltonian can be written as
\begin{equation}
    \hat{H}^\sigma
    =
    \hat{H}^\sigma_\mathrm{DFT}
    +
    \hat{V}^\sigma_\mathrm{HSE} \,,
    \label{eq:Hamiltonian_HSE}
\end{equation}
where $\hat{H}^\sigma_\mathrm{DFT}$ is the PBE Kohn-Sham Hamiltonian, including the full Perdew-Burke-Ernzerhof (PBE) xc potential~\cite{Perdew:1996}. The HSE contribution $\hat{V}^\sigma_\mathrm{HSE}$ replaces a fraction $\alpha$ of the short-range PBE exchange potential with the corresponding short-range Fock exchange potential~\cite{Ivady:2014}:
\begin{equation}
    \hat{V}^\sigma_\mathrm{HSE}
    =
    \alpha
    \left(
    \hat{V}^\sigma_\mathrm{Fock,sr}
    -
    \hat{V}^\sigma_\mathrm{PBEx,sr}
    \right) \,,
    \label{eq:pot_EXX}
\end{equation}
where $\hat{V}^\sigma_\mathrm{PBEx,sr}$ is the short-range part of the local PBE exchange potential, while $\hat{V}^\sigma_\mathrm{Fock,sr}$ is the corresponding nonlocal short-range Fock exchange potential. In the coordinate representation, the latter is given by
\begin{equation}
    V^\sigma_\mathrm{Fock,sr}(\mathbf{r},\mathbf{r}')
    =
    -
    \sum_{v,\mathbf{k}}
    f^\sigma_{v,\mathbf{k}}
    \frac{
        \mathrm{erfc}\!\left(\eta|\mathbf{r}-\mathbf{r}'|\right)
    }{
        |\mathbf{r}-\mathbf{r}'|
    }
    \psi^\sigma_{v,\mathbf{k}}(\mathbf{r})
    \psi^{\sigma *}_{v,\mathbf{k}}(\mathbf{r}') \,,
\end{equation}
where $\mathrm{erfc}$ is the complementary error function. The parameter $\eta$ controls the range separation between the short- and long-range parts of the Coulomb interaction. For HSE06, it is fixed to $\eta = 0.106\,a_0^{-1}$, where $a_0$ is the Bohr radius~\cite{Heyd:2006}. The fraction of short-range Fock exchange is fixed to $\alpha=0.25$ in the standard HSE06 functional.

Although $\alpha=0.25$ is often used in the HSE calculations for solids, its optimal value depends on the material and its electronic screening, in much the same way that the Hubbard parameter $U$ in DFT+$U$ depends on the material~\cite{Bastonero:2025}. In particular, the fraction of exact exchange is approximately related to the inverse of the high-frequency electronic dielectric constant, $\alpha \sim 1/\epsilon_\infty$~\cite{Skone:2014}. Since, to the best of our knowledge, the experimental value of $\epsilon_\infty$ has not been reported for DyFeO$_3$, we vary $\alpha$ and examine the resulting changes in the electronic structure. A first-principles determination of $\alpha$ based on the relation $\alpha \sim 1/\epsilon_\infty$ provides an interesting alternative~\cite{Riemelmoser:2026}, but is beyond the scope of the present work.

\subsection{Computational details}
\label{sec:comput_details}

All calculations are performed within the plane-wave pseudopotential framework as implemented in the \textsc{Quantum ESPRESSO} (QE) distribution~\cite{Giannozzi:2009, Giannozzi:2017, Giannozzi:2020}. We employ GGA with the PBE parametrization~\cite{Perdew:1996}. For Fe and O we use norm-conserving scalar-relativistic pseudopotentials from the PseudoDojo library (ONCVPSP v0.5, ``standard'' accuracy)~\cite{vanSetten:2018}. For Dy, we generate a norm-conserving scalar-relativistic pseudopotential with the $4f$ electrons explicitly included in the valence manifold ($5s^2 5p^6 4f^9 6s^2 5d^1$ electronic configuration). Spin-orbit coupling (SOC) is neglected throughout, both to simplify the analysis of the electronic structure and because it is not currently supported in the XANES calculations. All DFT calculations are performed within the spin-polarized, collinear framework, using the simplified magnetic configuration shown in Fig.~\ref{fig:Crystal_structure}(b). The calculations are performed at the experimental geometry, using the experimental lattice parameters and atomic positions of orthorhombic DyFeO$_3$ with space group $Pbnm$, as determined from powder X-ray diffraction (XRD) measurements~\cite{Biswas:2022}. The Kohn-Sham wavefunctions and the charge density and potentials are expanded in plane waves up to kinetic-energy cutoffs of 100 and 400~Ry, respectively. For ground-state calculations, the first Brillouin zone is sampled using a uniform $\Gamma$-centered $\mathbf{k}$-point mesh of $8\times8\times6$. The PDOS is calculated using a Gaussian smearing function with a broadening of $2\times10^{-3}$~Ry.

For PBE+$U$ calculations, the on-site Hubbard parameters $U$ for the Fe-$3d$ and Dy-$4f$ states are determined self-consistently from first principles using DFPT at the fixed experimental geometry~\cite{timrov2018hubbard,Timrov:2021}. The calculations are performed using the \textsc{HP} code~\cite{TIMROV-HP-2022} of the QE distribution~\cite{Giannozzi:2009,Giannozzi:2017,Giannozzi:2020}. As Hubbard projector functions, we employ L\"owdin-orthogonalized atomic orbitals~\cite{Lowdin:1950,Mayer:2002,Timrov:2020b}. In this scheme, the L\"owdin orthogonalization is performed over the complete set of atomic orbitals provided by the pseudopotentials, which are all charge neutral.
The Brillouin zone is sampled using uniform $\Gamma$-centered $\mathbf{k}$- and $\mathbf{q}$-point meshes of $4\times4\times3$ and $2\times2\times2$, respectively. The Hubbard parameters are converged to an accuracy of approximately $0.01$\,eV. Following the self-consistent procedure described in Ref.~\cite{Timrov:2021}, the Hubbard parameters are iteratively recalculated at the fixed experimental geometry until self-consistency is reached. The resulting values of $U$ are reported in Table~\ref{tab:Hub_param}.

\begin{table}[t]
    \renewcommand{\arraystretch}{1.3}
    \centering
    \caption{Comparison of the Hubbard $U$ parameters (in eV) used for the Fe-$3d$ and Dy-$4f$ states in DyFeO$_3$ in previous studies and in the present work. The previous studies determined the Hubbard parameters empirically, whereas the values reported in this work are obtained from first principles (see Sec.~\ref{sec:comput_details}). All studies employ PBE+$U$ as the underlying xc functional. Since the Hubbard projectors differ between the codes, the numerical values of $U$ are not directly transferable. Consequently, the comparison of $U$ values across different studies should be regarded only as an approximate measure of their relative magnitude.}
    \begin{tabular}{lllcc} 
    \hline\hline
     Study                    & Code   & $U$ determination      & Fe-$3d$ & Dy-$4f$ \\
     \hline
     Ref.~\cite{Cui:2024}     & VASP   & Empirical          &  4.5    & 6.0     \\
     Ref.~\cite{Stroppa:2010} & VASP   & Empirical          &  3.0    & 4.0     \\
     Ref.~\cite{Ameri:2021}   & Wien2k & Empirical          &  4.0    & 7.0     \\
     This work                & QE     & \textit{Ab initio} &  4.7    & 5.2     \\
    \hline\hline
    \end{tabular}
    \label{tab:Hub_param}
\end{table}

For the HSE calculations~\cite{Heyd:2003,Heyd:2006}, we use kinetic-energy cutoffs of 100~Ry for the Kohn-Sham wavefunctions, 400~Ry for the charge density and potentials, and 100~Ry for the exact-exchange contribution. The Brillouin zone is sampled using a $\Gamma$-centered $\mathbf{k}$-point mesh of $4\times4\times3$, and a $4\times4\times3$ $\mathbf{q}$-point mesh is used for the evaluation of the exact-exchange term. The $\mathbf{q}\rightarrow0$ limit is treated using the Gygi--Baldereschi scheme~\cite{Gygi1986}.

XANES calculations are performed using the \textsc{XSpectra} code of QE. The spectra are calculated according to Eq.~\eqref{eq:cross_section} on top of the DFT(+$U$) ground state using the Lanczos recursive algorithm~\cite{Taillefumier:2002,Gougoussis:2009}. Following Ref.~\cite{Timrov:2020c}, we neglect the core hole associated with the O($1s$) electron promoted to the conduction manifold. In preliminary calculations for DyFeO$_3$, including the core hole was found to substantially dampen the first two peaks of the O $K$-edge spectrum and thereby worsen the agreement with experiment. We therefore neglect core-hole effects in the results presented here. All XANES spectra are calculated within the dipole approximation, neglecting quadrupole and higher-order contributions. To account for the powder character of the experimental samples, the XANES intensity is averaged over three mutually perpendicular polarizations of the incident photon beam, taken along the Cartesian $x$, $y$, and $z$ directions. For each polarization, the spectra are calculated for the two inequivalent O sites in DyFeO$_3$, and the resulting spectra are subsequently averaged with the proper weights. The calculated powder-averaged spectra are then compared with the experimental powder spectra. A uniform $\Gamma$-centered $\mathbf{k}$-point mesh of $8\times8\times6$ is used for the XANES calculations. The calculated spectra are broadened using a Lorentzian function with an energy-dependent broadening parameter~\cite{Bunau:2013}. The broadening increases from 0.16\,eV, corresponding to the O($1s$) core-hole lifetime broadening~\cite{Sankari:2003}, to a maximum value of 2\,eV. Its energy dependence follows an arctangent-type function with an inflection point located 12\,eV above the valence-band maximum~\cite{Bunau:2013}. All calculated spectra are rigidly shifted in energy to align the position of the first excitation peak with that of the experimental spectrum.


\subsection{Experimental details}

DyFeO$_3$ powder was synthesized by a conventional solid-state reaction method. Stoichiometric amounts of dried Fe$_2$O$_3$ and Dy$_2$O$_3$ powders, both with a purity of 99.99\%, were thoroughly mixed and ground, pressed into a pellet, and subsequently sintered. The pellet was first calcined in air at 1100$^{\circ}$C for 24~h. It was then re-ground, pressed, and sintered again in air at 1300$^{\circ}$C for 24~h, followed by cooling at a rate of approximately 1$^{\circ}$C/min to ensure the desired oxygen stoichiometry. Scanning electron microscopy confirmed a grain size of several micrometers, while Rutherford backscattering measurements confirmed the chemical composition of the resulting powder~\cite{Doebeli:2008}.

Powder XANES measurements were performed at the SIM beamline of the Swiss Light Source~\cite{windsor:2014, flechsig:2010}. The O $K$-edge spectra were collected at room temperature with the incident X-ray beam at an angle of 45$^\circ$ to the sample surface. Measurements were performed in both total electron yield (TEY) and total fluorescence yield (TFY) modes, providing sensitivity to different probing depths~\cite{abbate:1992, asakura:2016}. The measured spectra were first normalized to the incident photon flux recorded using a clean gold mesh positioned upstream of the experimental chamber. The resulting spectra were then normalized using the absorption-edge jump determined from spectral regions well below and above the O $K$-edge. For qualitative analysis, a polynomial background was fitted to and subtracted from the normalized spectra. All XANES data were processed using the \textsc{Athena} software package~\cite{ravel:2005}.  
 
\section{Results and Discussion}
\label{sec:Results_and_Discussion}

A reliable description of the electronic structure of DyFeO$_3$ is an essential prerequisite for interpreting its O $K$-edge XANES spectra. We therefore first assess the electronic structure with various xc functionals of DFT. We begin with the semilocal PBE functional and subsequently consider PBE+$U$, with the Hubbard parameters obtained \textit{ab initio} using DFPT. We then further examine the electronic structure using the HSE hybrid functional. By systematically varying the $\alpha$ mixing parameter, we identify the value that provides the best overall agreement with selected experimental observables. The resulting HSE electronic structure is then used to calibrate the parameters of the OR-PBE+$U$ approach employed in the subsequent XANES calculations. This hierarchy of calculations allows us to assess systematically how the treatment of electronic interactions affects the electronic states relevant to the O $K$-edge spectra.

\section{Electronic structure}

\subsection{PBE and PBE+$U$}
\label{sec:PBE_andPBE+U}

\begin{figure*}[t]
    \centering
    \begin{subfigure}{0.49\textwidth}
        \centering
        \includegraphics[width=\linewidth]{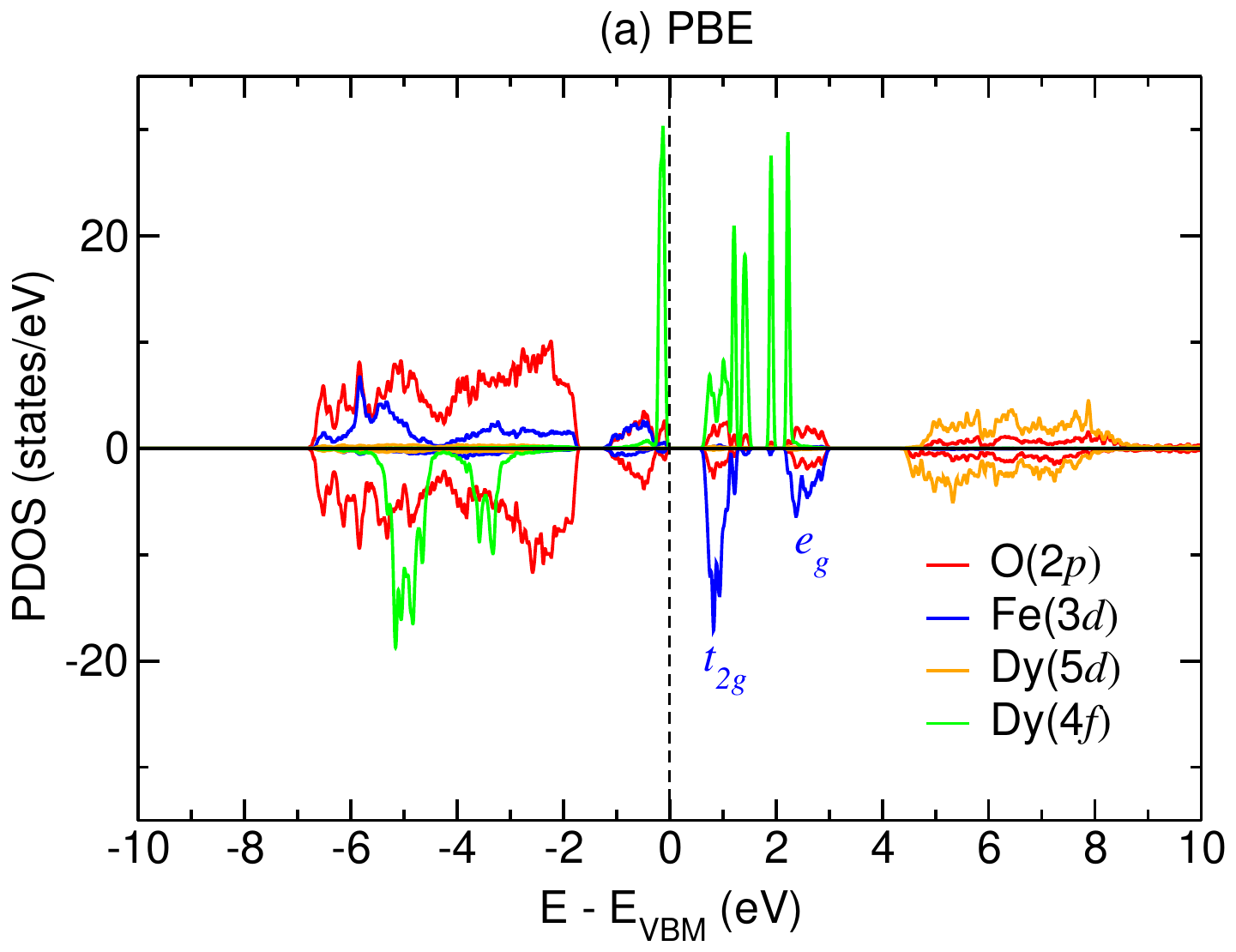}
    \end{subfigure}
    \hfill
    \begin{subfigure}{0.49\textwidth}
        \centering
        \includegraphics[width=\linewidth]{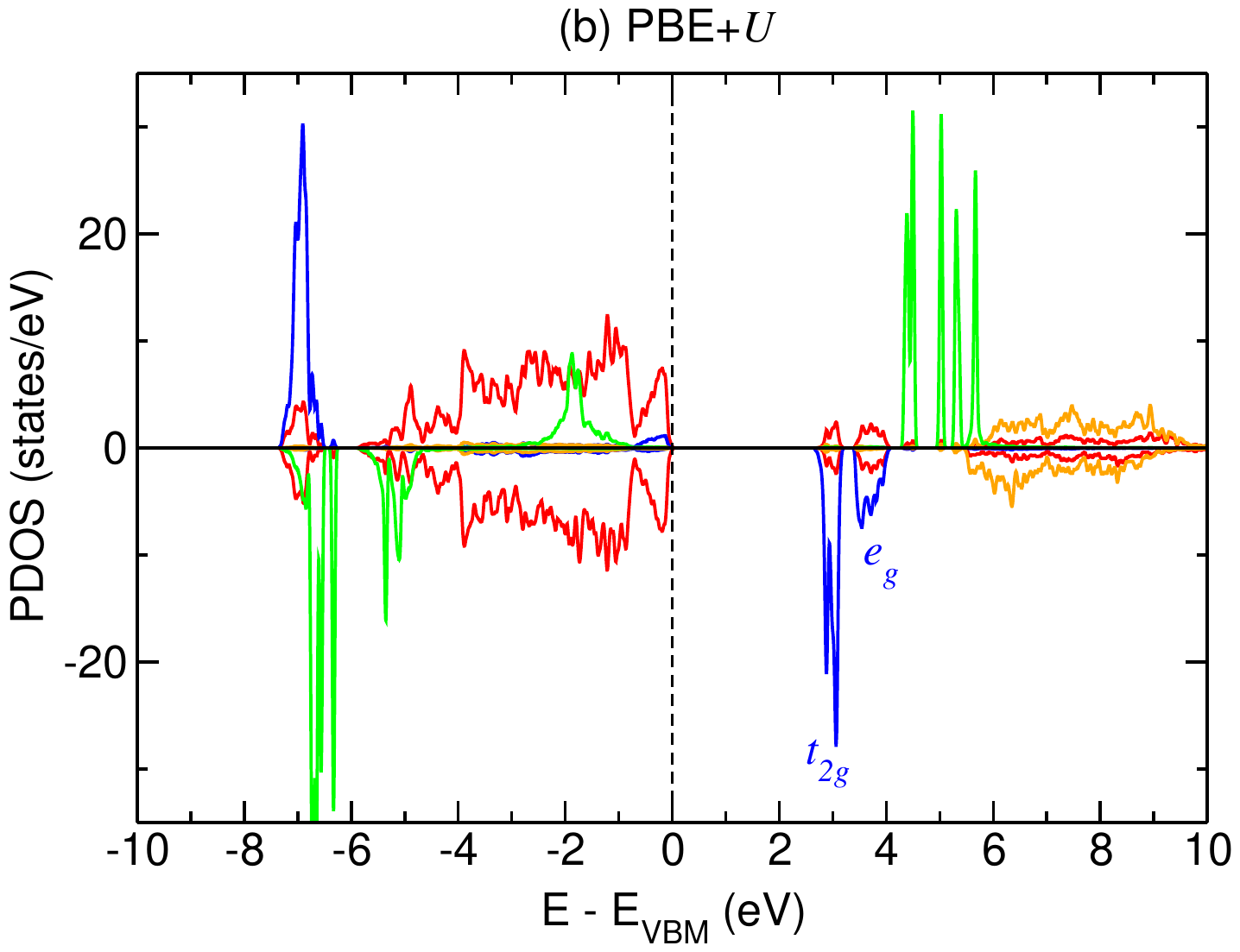}
    \end{subfigure}
    \caption{PDOS of DyFeO$_3$ calculated using (a) PBE and (b) PBE+$U$ with the \emph{ab initio} Hubbard parameters for the Fe-$3d$ and Dy-$4f$ states given in Table~\ref{tab:Hub_param}. The electronic structure is calculated for the collinear antiferromagnetic configuration shown in Fig.~\ref{fig:Crystal_structure}(b). The upper and lower halves of each panel show the spin-up and spin-down components, respectively, with the latter plotted with negative sign. For each chemical species, only one of the two symmetry-equivalent magnetic sublattices is shown; the corresponding states of the oppositely polarized sublattice are omitted for clarity. The energy zero is set to the VBM.}
    \label{fig:PDOS_PBE_vs_PBE+U}
\end{figure*}

We first consider the electronic structure obtained within PBE. Figure~\ref{fig:PDOS_PBE_vs_PBE+U}(a) presents the corresponding PDOS, resolved into the O-$2p$, Fe-$3d$, Dy-$4f$, and Dy-$5d$ contributions. PBE predicts DyFeO$_3$ to be insulating, consistent with experiment, but substantially underestimates the fundamental band gap, yielding $0.6$\,eV. This value is considerably smaller than the experimental optical gap of approximately $2.1$\,eV~\cite{Tarek:2024,NoorjahanBegum:2025,Vani:2026}, illustrating the well-known tendency of semilocal xc functionals to underestimate band gaps~\cite{Perdew:2017}. The comparison should nevertheless be made with caution: the optical gap is not, in general, identical to the fundamental quasiparticle gap, because electron-hole interactions can give rise to excitonic effects and thereby shift the optical absorption onset to lower energies. Since PBE does not account for such effects, the experimental optical gap should be regarded here primarily as a useful reference for assessing the magnitude of the PBE error rather than as a direct measure of the fundamental gap. Since the PBE fundamental gap is already smaller than the experimental optical gap, accounting for excitonic effects, which would further reduce the gap, would only increase the discrepancy with experiment.

The orbital character of the states near the gap provides further insight into the limitations of the PBE description. As shown in Fig.~\ref{fig:PDOS_PBE_vs_PBE+U}(a), the valence-band maximum (VBM) has a pronounced Dy-$4f$ contribution, whereas the conduction-band minimum (CBM) contains predominantly Fe-$3d$ and Dy-$4f$ character, with additional O-$2p$ hybridization. The unoccupied Fe-$3d$ states between approximately $1$ and $3$\,eV exhibit a clear crystal-field splitting into the $t_{2g}$ and $e_g$ manifolds (strictly speaking, these labels are approximate for the distorted FeO$_6$ octahedra, but we retain this nomenclature for simplicity). Within PBE, the corresponding splitting is approximately $1.7$\,eV. This energy separation is of particular importance for the interpretation of the O $K$-edge XANES spectra because transitions from O-$1s$ states probe unoccupied states with O-$2p$ character, and the Fe-$3d$ crystal-field splitting can therefore be reflected in the spectral structure through the hybridization between the Fe-$3d$ and O-$2p$ states~\cite{Frati:2020, Timrov:2020c}. At higher energies, between approximately $5$ and $9$\,eV, the unoccupied manifold acquires substantial Dy-$5d$ character. In contrast to the relatively narrow Dy-$4f$ states, the more spatially extended Dy-$5d$ states form a broader manifold and consequently exhibit stronger energetic overlap with the O-$2p$ states. The occupied manifold, extending approximately from $-7$ to $-1.5$\,eV relative to the VBM, contains a substantial O-$2p$ contribution that overlaps strongly with both Fe-$3d$ and Dy-$4f$ states. Thus, the PBE electronic structure is characterized not only by a severely underestimated band gap but also by a pronounced sensitivity of the relative positions of states to the localization of the Fe-$3d$ and Dy-$4f$ electrons. This motivates the use of a Hubbard correction on top of semilocal PBE.

We therefore next consider PBE+$U$, following the general approach adopted in previous computational studies of DyFeO$_3$~\cite{Stroppa:2010, Ameri:2021, Cui:2024}. A key distinction of the present work is that the Hubbard parameters for the Fe-$3d$ and Dy-$4f$ states are obtained \textit{ab initio}, rather than chosen empirically. The resulting values are of similar magnitude to those employed in previous studies, although differences of nearly $2$\,eV occur for individual parameters (see Table~\ref{tab:Hub_param}). Such numerical comparisons, however, should be interpreted with caution. The value of an effective Hubbard parameter is not a universal material constant: it depends on the choice of  Hubbard projectors and other computational details~\cite{Timrov:2018}. Consequently, comparisons of $U$ values obtained using different computational frameworks can provide useful insight into their overall magnitude, but relatively small differences should be interpreted with care, given their dependence on the computational details and the method used to determine $U$~\cite{Carta:2026, Yang:2026}. Nevertheless, the overall similarity in the magnitude of the parameters in Table~\ref{tab:Hub_param} provides a useful point of reference. At the level of the electronic structure, our PBE+$U$ PDOS is in very good qualitative agreement with the results reported in Refs.~\cite{Ameri:2021,Cui:2024}, including the overall distribution and orbital character of the occupied and unoccupied states.

The PBE+$U$ PDOS, shown in Fig.~\ref{fig:PDOS_PBE_vs_PBE+U}(b), exhibits several substantial changes relative to PBE. Most notably, the fundamental gap increases to approximately $2.6$\,eV, bringing the calculated gap much closer to the experimental optical gap. As noted above, the two quantities are not strictly equivalent because excitonic effects can lower the optical absorption onset relative to the fundamental gap. Nevertheless, the substantially improved gap provides clear evidence that the Hubbard correction remedies an important deficiency of the PBE functional~\cite{KirchnerHall:2021}. The improvement in the gap is accompanied by a significant redistribution of the unoccupied Fe-$3d$ states. In particular, the separation between the $t_{2g}$ and $e_g$ manifolds is reduced from approximately $1.7$\,eV within PBE to only about $0.7$\,eV in PBE+$U$. This reduction is important for the interpretation of the O $K$-edge spectra, where the relative energy of the Fe-$3d$-derived states directly influences the spectral structure. A similar reduction of the Fe-$3d$ crystal-field splitting upon inclusion of a Hubbard correction has also been reported for LaFeO$_3$~\cite{Timrov:2020c} and YFeO$_3$~\cite{Stoeffler:2017}. Thus, although PBE+$U$ substantially improves the fundamental band gap, it does not necessarily provide a uniformly improved description of all electronic states relevant to the XANES spectra. The Hubbard correction also produces pronounced changes in the Dy-derived states. The unoccupied Dy-$4f$ states are shifted away from the band-edge region and are redistributed predominantly over the approximately $4-6$\,eV energy range, where they touch the Dy-$5d$ manifold. Although no explicit Hubbard correction is applied to the Dy-$5d$ states, their spectral position is nevertheless modified indirectly through the self-consistent rearrangement of the electronic structure and the accompanying changes in hybridization with the Hubbard-corrected Fe-$3d$ and Dy-$4f$ states. This illustrates that the effect of the Hubbard correction extends beyond the explicitly corrected orbitals. The occupied manifold is modified even more substantially. In contrast to PBE, where the VBM has a strong Dy-$4f$ character, the VBM in PBE+$U$ is dominated by O-$2p$ states. The occupied Fe-$3d$ states become considerably more localized and are shifted to lower energies, with substantial spectral weight appearing around $-7$\,eV. The occupied Dy-$4f$ states also undergo a pronounced redistribution. In the spin-up channel, the Dy-$4f$ spectral weight that was dominant in the VBM within PBE is strongly suppressed, broadened, and shifted toward approximately $-2$\,eV. In the spin-down channel, the occupied Dy-$4f$ states are even more localized than in PBE and are shifted to lower energies, contributing predominantly in the range from approximately $-5$ to $-7$\,eV. These changes substantially alter the orbital character and hybridization of the occupied valence manifold relative to the PBE description.

The resulting PBE+$U$ electronic structure therefore presents a nontrivial trade-off. The Hubbard correction yields a much more realistic fundamental band gap and localizes the Fe-$3d$ and Dy-$4f$ states, but at the same time it substantially modifies the Fe-$3d$ crystal-field splitting and the detailed distribution of spectral weight across the occupied and unoccupied manifolds. Whether this redistribution provides a more accurate description of the experimentally relevant electronic structure cannot be established from the band gap alone. Direct comparison with photoemission measurements of the occupied states would provide an important additional test, but such an analysis is beyond the scope of the present work. Instead, we use the O $K$-edge XANES response as an independent probe of the unoccupied electronic structure. Before turning to the XANES calculations, however, it is useful to establish whether the qualitative electronic structure obtained with PBE+$U$ is consistent with predictions from other advanced xc functionals such as HSE. We therefore examine next the electronic structure using the HSE hybrid functional and assess systematically its dependence on the exact-exchange mixing parameter $\alpha$.

\subsection{HSE}

\begin{figure*}[t]
    \centering
    \includegraphics[width=0.85\linewidth]{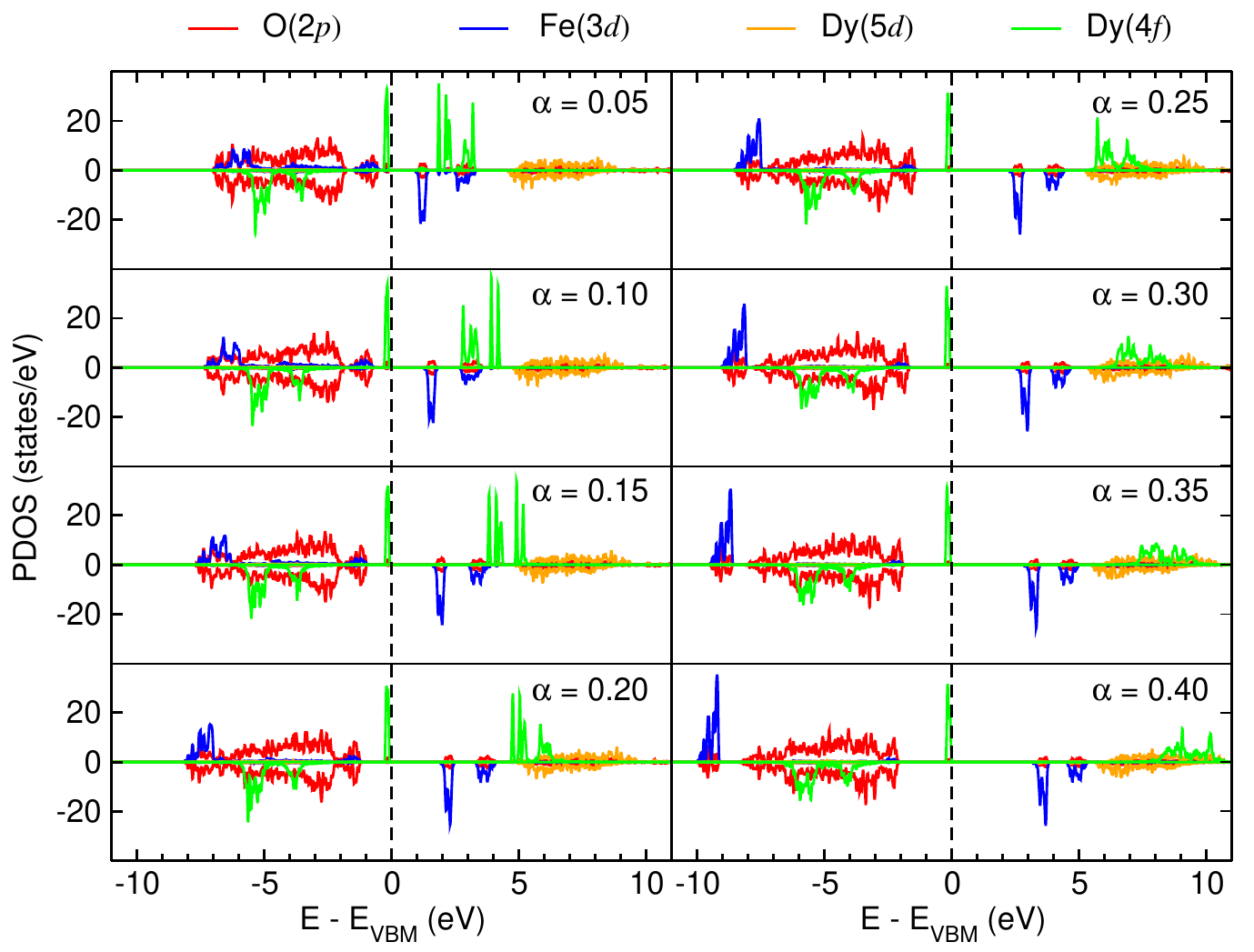}
    \caption{PDOS of DyFeO$_3$ calculated using the HSE hybrid functional for exact-exchange mixing parameters $\alpha$ ranging from 0.05 to 0.40. The electronic structure is calculated for the collinear antiferromagnetic configuration shown in Fig.~\ref{fig:Crystal_structure}(b). The upper and lower halves of each panel show the spin-up and spin-down components, respectively, with the latter plotted with negative sign. For each chemical species, only one of the two symmetry-equivalent magnetic sublattices is shown; the corresponding states of the oppositely polarized sublattice are omitted for clarity. The energy zero is set to the VBM.}
    \label{fig:PDOS_HSE}
\end{figure*}

Figure~\ref{fig:PDOS_HSE} shows the PDOS of DyFeO$_3$ calculated using the HSE hybrid functional for exact-exchange mixing parameters $\alpha$ ranging from 0.05 to 0.40. The conventional HSE06 functional corresponds to $\alpha=0.25$~\cite{Heyd:2003,Heyd:2006}. At the lowest value considered, $\alpha=0.05$, the electronic structure remains qualitatively close to the PBE result ($\alpha=0$), shown in Fig.~\ref{fig:PDOS_PBE_vs_PBE+U}(a). Increasing $\alpha$ progressively modifies both the band gap and the energetic distribution of the orbital-projected states, revealing systematic trends that are useful for assessing the suitability of the hybrid-functional calculations.

Before discussing these trends, it is useful to emphasize an important distinction between the treatment of exchange in HSE and the Hubbard correction employed within DFT+$U$. In HSE, the same fraction $\alpha$ of screened Fock exchange is applied to all occupied and unoccupied electronic states, irrespective of their orbital character. In DyFeO$_3$, however, the relevant $s$, $p$, $d$, and $f$ states differ substantially in their spatial localization and degree of hybridization. Consequently, the value of $\alpha$ that provides an appropriate correction for the relatively delocalized $s$ and $p$ states need not be optimal for the more localized transition-metal $d$ and rare-earth $f$ states~\cite{Ivady:2014}. The present calculations therefore probe an effective, system-wide value of $\alpha$, applied uniformly to all electronic states. This feature can be contrasted with the orbital-dependent nature of DFT+$U$, where separate Hubbard parameters can be assigned to different states, as illustrated here by the Fe-$3d$ and Dy-$4f$ states (see Table~\ref{tab:Hub_param}). In this context, it is worth noting the approach proposed by Ivády \textit{et al.}~\cite{Ivady:2014}, in which a hybrid-functional correction is combined with additional Hubbard-like corrections for more localized states. In such a scheme, a common exact-exchange contribution can account for the self-interaction errors associated with more delocalized states, while additional orbital-dependent Hubbard-like corrections provide a stronger treatment of the localized manifolds. We do not pursue such an approach here though, leaving it for future studies. Instead, we perform a systematic variation of a global, effective $\alpha$ parameter, which provides a controlled way of examining how a uniform treatment of exact exchange modifies the different orbital manifolds and, ultimately, which value of $\alpha$ yields an electronic structure most consistent with the available experimental data.

We first consider the evolution of the fundamental band gap. At $\alpha=0.05$, the calculated gap is approximately $1.1$~eV, increasing continuously to approximately $3.2$\,eV at $\alpha=0.40$. For HSE06, we obtain a fundamental band gap of approximately $2.3$\,eV. Thus, the band gap increases systematically with the amount of exact exchange, and the HSE06 value is in good agreement with the experimental optical gap~\cite{Tarek:2024, NoorjahanBegum:2025, Vani:2026}. As discussed above, this comparison is not strictly quantitative because the fundamental and optical gaps are distinct quantities. Nevertheless, the evolution of the calculated gap with $\alpha$ provides a useful criterion for assessing the physically relevant range of the hybrid-functional parameter $\alpha$.

The unoccupied states exhibit a clear and systematic response to increasing $\alpha$. The Fe-$3d$ and Dy-$4f$ states progressively shift toward higher energies. Interestingly, the Fe-$3d$ crystal-field splitting is comparatively insensitive to $\alpha$. The separation between the $t_{2g}$ and $e_g$ manifolds decreases only gradually, from approximately $1.6$\,eV at $\alpha=0.05$ to approximately $1.3$\,eV at $\alpha=0.40$, despite the substantial upward shift of the Fe-$3d$ manifold as a whole. This behavior contrasts with the much stronger dependence of the Dy-$4f$ states on $\alpha$. The unoccupied Dy-$4f$ spectral weight moves rapidly toward higher energies and becomes progressively broader, resulting in a reduction of its peak intensity. The Dy-$5d$ states, in contrast, are comparatively insensitive to $\alpha$, apart from a modest shift toward higher energies.

The occupied states respond differently to the increasing exact-exchange fraction $\alpha$. The Fe-$3d$ states become progressively more localized and shift toward lower energies as $\alpha$ increases, reaching energies as low as approximately $-9.5$\,eV for $\alpha=0.40$. The behavior of the occupied Dy-$4f$ states is particularly noteworthy. In the spin-up channel, these states remain sharply localized and their position changes only weakly with $\alpha$. They therefore continue to contribute prominently to the VBM throughout the entire range of $\alpha$ considered. This behavior differs markedly from the PBE+$U$ result [Fig.~\ref{fig:PDOS_PBE_vs_PBE+U}(b)], where the Hubbard correction strongly suppresses the Dy-$4f$ spectral weight at the VBM and shifts these states to lower energies. The occupied Dy-$4f$ states in the spin-down channel are also comparatively insensitive to $\alpha$. Their spectral weight is somewhat reduced and shifts modestly toward lower energies, but the effect is substantially weaker than the pronounced localization and downward shift obtained within PBE+$U$.

These results demonstrate that HSE and PBE+$U$ do not produce equivalent corrections to the PBE electronic structure. Although both approaches increase the fundamental band gap and modify the positions of the Fe-$3d$ and Dy-$4f$ states, they redistribute the spectral weight in qualitatively different ways. This distinction becomes particularly apparent when comparing HSE06 with the PBE+$U$ electronic structure. At $\alpha=0.25$, the overall distribution of the occupied and unoccupied states is qualitatively similar to that obtained with PBE+$U$, but several important differences remain. Most notably, the VBM in HSE06 retains a pronounced Dy-$4f$ character, whereas in PBE+$U$ the occupied Dy-$4f$ states are strongly suppressed and shifted to lower energies, leaving an O-$2p$-dominated VBM. The fundamental gap is also somewhat smaller in HSE06, approximately $2.3$\,eV compared with $2.6$\,eV in PBE+$U$, and is consequently closer to the experimental optical gap.
An additional important distinction concerns the Fe-$3d$ crystal-field splitting. For HSE06, the separation between the unoccupied $t_{2g}$ and $e_g$ manifolds is approximately $1.4$\,eV, substantially larger than the $\sim0.7$\,eV obtained with PBE+$U$. As discussed in Sec.~\ref{sec:XANES}, the former value is also in much better agreement with the experimentally inferred crystal-field splitting. Thus, although PBE+$U$ provides a substantial improvement over PBE, HSE06 provides a more balanced description of the electronic energy scales that are particularly relevant to the O $K$-edge XANES spectra. In particular, its simultaneous description of the fundamental gap and the Fe-$3d$ crystal-field splitting makes HSE06 a promising reference electronic structure for the subsequent analysis.

A remaining issue is the choice of optimal $\alpha$. As was discussed earlier, the exact-exchange fraction is material dependent and is related to the electronic screening, with $\alpha$ often interpreted approximately in terms of the inverse high-frequency dielectric constant, $\alpha\sim1/\epsilon_{\infty}$ (see Sec.~\ref{sec:comput_approach}). Since $\epsilon_{\infty}$ is not available for DyFeO$_3$, we cannot determine $\alpha$ from this relation. We therefore adopt the conventional HSE06 value, $\alpha=0.25$, as a reference. This choice is further supported by the satisfactory agreement of the resulting electronic structure with the available experimental data, particularly the fundamental band gap and the Fe-$3d$ crystal-field splitting. 

However, XANES calculations based on HSE are presently not feasible within the computational framework employed here because the required implementation is not available. We therefore use the HSE06 results as a target electronic structure and investigate how the OR-PBE+$U$ framework can be empirically parametrized to reproduce its key features. In the following section, we accordingly tune the orbital-resolved $U$ parameters of OR-PBE+$U$ with the aim of obtaining an electronic structure that closely reproduces the HSE06 reference PDOS while retaining the computational framework required for the subsequent XANES simulations.

\subsection{Orbital-resolved PBE+$U$}
\label{subsec:or-pbe-u}

\begin{figure*}[t]
    \centering
    \begin{subfigure}{0.49\textwidth}
        \centering
        \includegraphics[width=\linewidth]{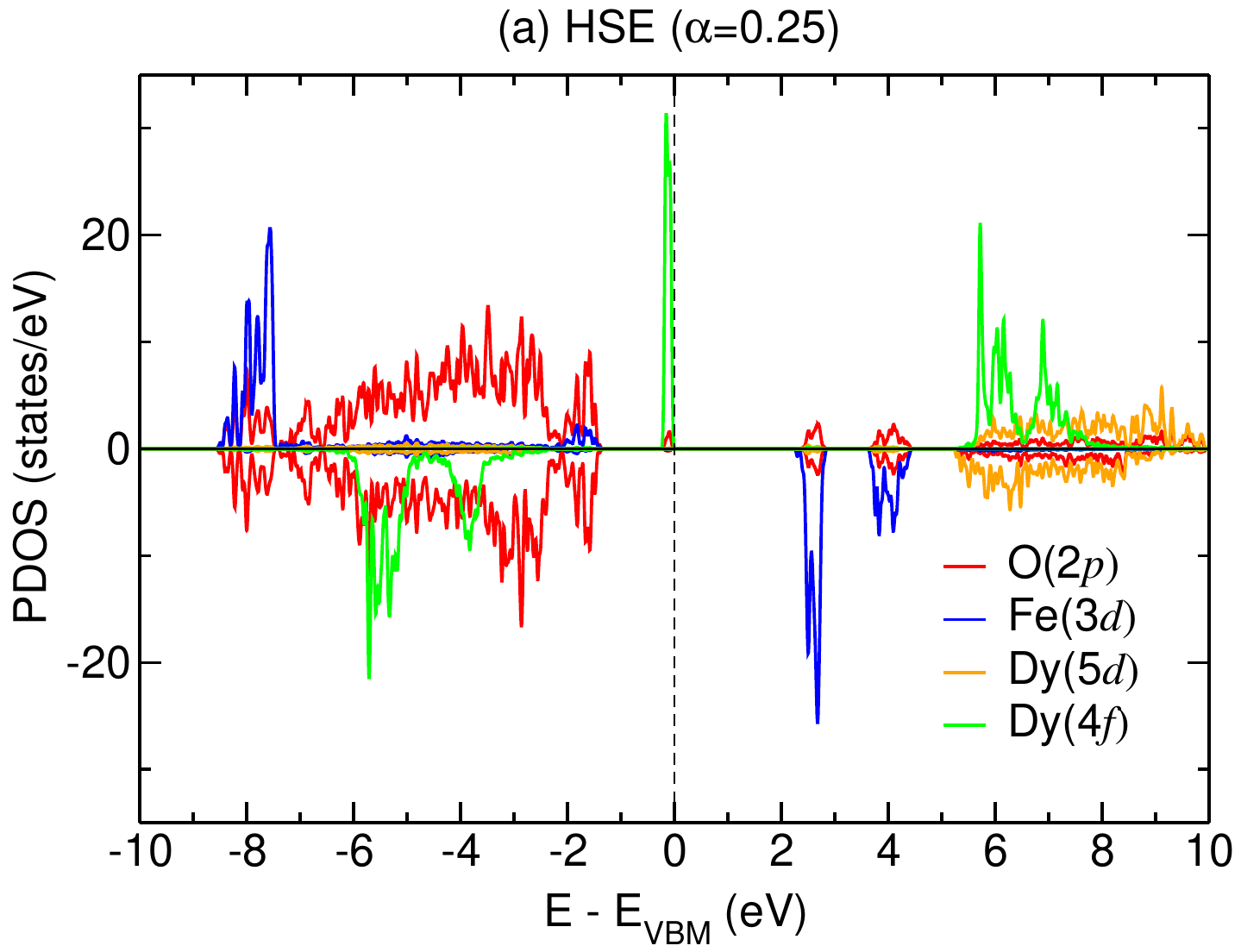}
    \end{subfigure}
    \hfill
    \begin{subfigure}{0.49\textwidth}
        \centering
        \includegraphics[width=\linewidth]{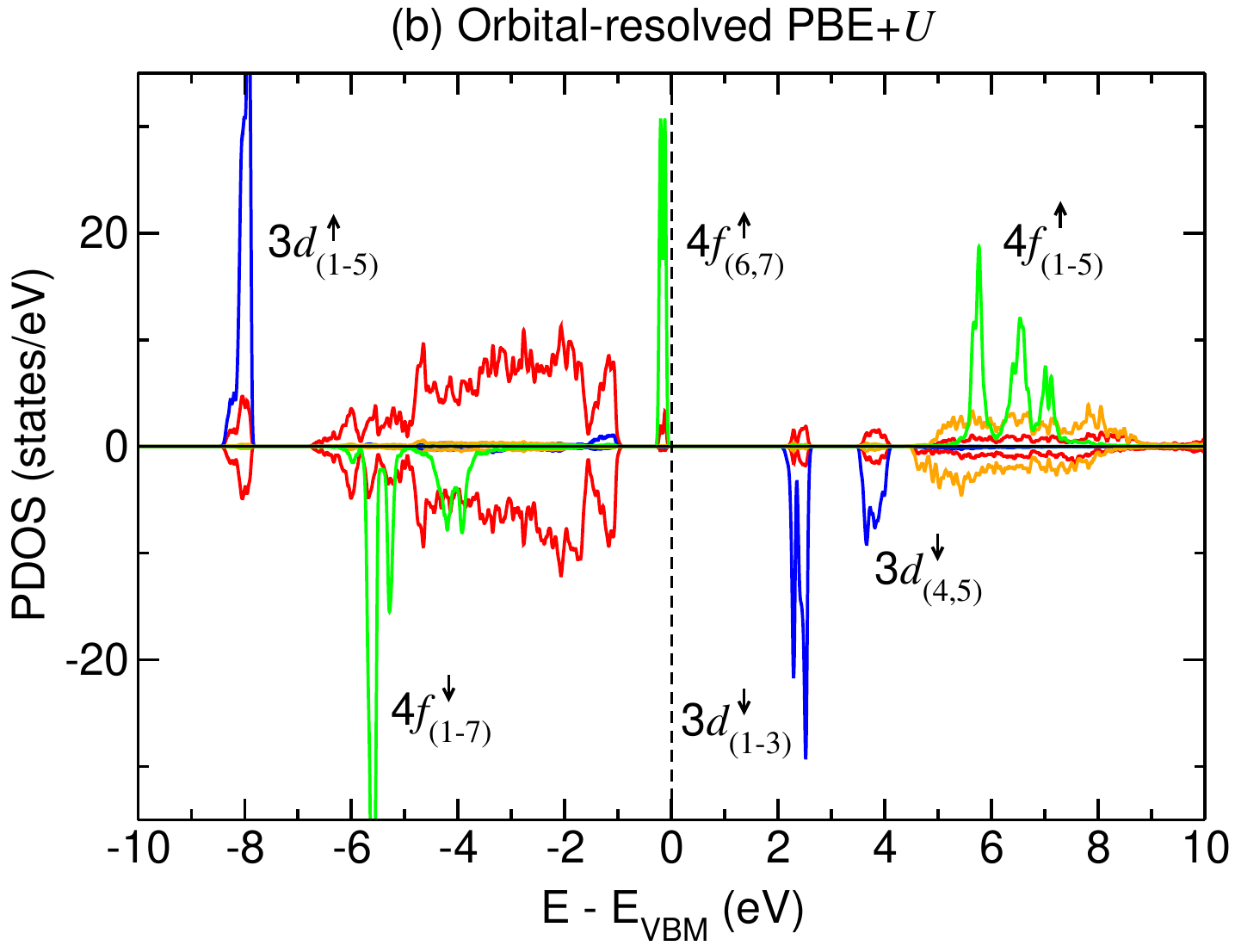}
    \end{subfigure}
    \caption{PDOS of DyFeO$_3$ calculated using (a) HSE with $\alpha = 0.25$ and (b) OR-PBE+$U$ with orbital-resolved Hubbard $U$ parameters listed in Table~\ref{tab:Hub_param_OR}. The electronic structure is calculated for the collinear antiferromagnetic configuration shown in Fig.~\ref{fig:Crystal_structure}(b). The upper and lower halves of each panel show the spin-up and spin-down components, respectively, with the latter plotted with negative sign. For each chemical species, only one of the two symmetry-equivalent magnetic sublattices is shown; the corresponding states of the oppositely polarized sublattice are omitted for clarity. The energy zero is set to the VBM.}
    \label{fig:PDOS_HSE_vs_OR-PBE+U}
\end{figure*}

As demonstrated in our previous study~\cite{Warda:2025}, introducing orbital resolution into the Hubbard correction provides additional flexibility for systems containing localized $d$ and $f$ electrons and can substantially improve the description of their electronic structure and related properties. Here, we apply this approach to DyFeO$_3$. Figure~\ref{fig:PDOS_HSE_vs_OR-PBE+U}(a) shows the PDOS calculated using HSE with $\alpha=0.25$, which we take as the reference electronic structure. Figure~\ref{fig:PDOS_HSE_vs_OR-PBE+U}(b) shows the corresponding PDOS obtained using OR-PBE+$U$, where the orbital-resolved Hubbard parameters were adjusted to reproduce the HSE06 electronic structure as closely as possible. The calibration of the orbital-resolved Hubbard parameters was performed by manually varying the individual $U$ values and selecting a set that approximately reproduces the positions and overall distribution of the Fe-$3d$ and Dy-$4f$ states in the HSE06 PDOS. Although more systematic parameter-optimization procedures, such as Bayesian optimization~\cite{Yu:2020}, could be employed, such an approach is beyond the scope of the present work. The resulting Hubbard parameters are summarized in Table~\ref{tab:Hub_param_OR}. In the Appendix, we also discuss our attempts to determine the orbital-resolved Hubbard parameters from first principles using DFPT.

\begin{table}[h!]
    \renewcommand{\arraystretch}{1.3}
    \centering
    \caption{Orbital-resolved Hubbard $U$ parameters used in OR-PBE+$U$, calibrated to reproduce the PDOS obtained from HSE06 with $\alpha=0.25$. The values shown correspond to one representative Fe ion and one representative Dy ion. The other Fe and Dy ions belonging to the oppositely polarized magnetic sublattice are symmetry equivalent to the respective ions shown here, with the $U$ values for the spin-up and spin-down channels interchanged.}
    \begin{tabular}{l|c|c|c|c|c|c} 
    \hline\hline
     \multirow{2}{*}{Manifold} & \multicolumn{3}{c|}{Fe-$3d$} & \multicolumn{3}{c}{Dy-$4f$} \\ \cline{2-7}
        & $3d^\uparrow_{(1-5)}$ & $3d^\downarrow_{(1-3)}$ & $3d^\downarrow_{(4-5)}$ & $4f^\uparrow_{(6,7)}$ & $4f^\uparrow_{(1-5)}$ & $4f^\downarrow_{(1-7)}$ \\ \hline
     $U$ (eV)    & 2.8 & 7.6 & 11.2 & 0.0 & 9.8 & 1.0  \\
    \hline\hline
    \end{tabular}
    \label{tab:Hub_param_OR}
\end{table}

A key result is that the agreement with HSE06 cannot be obtained with the same level of fidelity using a single Hubbard parameter for all Fe-$3d$ states and a single Hubbard parameter for all Dy-$4f$ states. Instead, resolving the Hubbard correction at the level of individual orbital submanifolds is essential. The relevant submanifolds can be identified by diagonalizing the atomic occupation matrix according to Eq.~\eqref{eq:occ_matrix_eigen}. The resulting occupation eigenstates are ordered according to their occupation eigenvalues, with occupied and unoccupied states corresponding to eigenvalues of one and zero, respectively. This provides a natural labeling of the orbital submanifolds used in the orbital-resolved Hubbard correction.

For the PDOS of the representative Fe ion shown in Fig.~\ref{fig:PDOS_HSE_vs_OR-PBE+U}(b), all five Fe-$3d$ states in the spin-up channel are occupied and are therefore treated as a single manifold, Fe-$3d^\uparrow_{(1-5)}$, with $U=2.8$\,eV. In the spin-down channel, all five Fe-$3d$ states are unoccupied and are split into two manifolds according to their crystal-field character: Fe-$3d^\downarrow_{(1-3)}$, corresponding to the $t_{2g}$ states, and Fe-$3d^\downarrow_{(4-5)}$, corresponding to the $e_g$ states. The corresponding Hubbard parameters are $U=7.6$\,eV and $11.2$\,eV, respectively. Thus, the unoccupied spin-down states require substantially larger corrections than the occupied spin-up states. Moreover, the distinct $U$ values assigned to the $t_{2g}$ and $e_g$ manifolds are important for reproducing the magnitude of the crystal-field splitting obtained within HSE06. This result illustrates that, for the Fe-$3d$ states, a single shell-averaged Hubbard parameter is insufficient to simultaneously reproduce the relative positions of the occupied and unoccupied manifolds.

The Dy-$4f$ states exhibit a qualitatively different behavior. In the spin-up channel, the two occupied states forming the VBM are labeled as Dy-$4f^\uparrow_{(6,7)}$. No Hubbard correction is required for these states, for which we set $U=0$. Applying a finite positive $U$ to these states would shift their spectral weight away from the VBM and alter their position relative to the HSE06 reference. The remaining five Dy-$4f$ states, Dy-$4f^\uparrow_{(1-5)}$, are unoccupied and require a substantially larger correction, $U=9.8$\,eV, to reproduce their position at higher energies. In the spin-down channel, all seven Dy-$4f$ states are occupied, forming the Dy-$4f^\downarrow_{(1-7)}$ manifold. Only a small correction, $U=1.0$\,eV, is required to reproduce the corresponding spectral distribution.

With these orbital-resolved parameters, the OR-PBE+$U$ PDOS reproduces the HSE06 electronic structure qualitatively well, including the positions and relative separation of the principal Fe-$3d$ and Dy-$4f$ manifolds. Quantitative differences remain, as expected, because the present calibration was designed to provide an approximate mapping of the HSE06 electronic structure onto the OR-PBE+$U$ framework rather than a fully optimized parameterization. Nevertheless, the comparison demonstrates that orbital resolution of the Hubbard parameters provides the flexibility required to reproduce the main features of the HSE06 PDOS, whereas shell-averaged $U$ values do not provide sufficient degrees of freedom for this purpose.

Some of the remaining discrepancies between the tuned OR-PBE+$U$ PDOS and the HSE06 reference occur for the Dy-$5d$ and O-$2p$ states. These differences originate from the fact that the Dy-$5d$ and O-$2p$ states are affected by the exchange-exchange correction in HSE06, whereas they are not explicitly corrected by a Hubbard term in the present OR-PBE+$U$ calculations. Their electronic structure is nevertheless modified indirectly through the self-consistent rearrangement of the electronic states induced by the orbital-resolved Hubbard corrections applied to the Fe-$3d$ and Dy-$4f$ states. In principle, an additional Hubbard correction could also be applied to the O-$2p$ states; however, we do not introduce such a correction here. For the Dy-$5d$ states, in contrast, applying an additional Hubbard correction is currently not possible within our implementation of OR-PBE+$U$, because the present implementation does not support simultaneous orbital-resolved Hubbard corrections for two different angular-momentum channels on the same atomic species. In particular, for Dy, the current OR-PBE+$U$ implementation cannot treat the $d$ and $f$ channels simultaneously. Consequently, the Dy-$5d$ states cannot be explicitly corrected while retaining the orbital-resolved Hubbard correction for the Dy-$4f$ states. The remaining differences in the Dy-$5d$ and O-$2p$ PDOS therefore reflect both the different treatments of electronic interactions within HSE06 and OR-PBE+$U$ and the present limitations of the Hubbard-correction scheme.

\section{XANES spectra}
\label{sec:XANES}

\begin{figure}[t]
    \centering
    \includegraphics[width=0.98\linewidth]{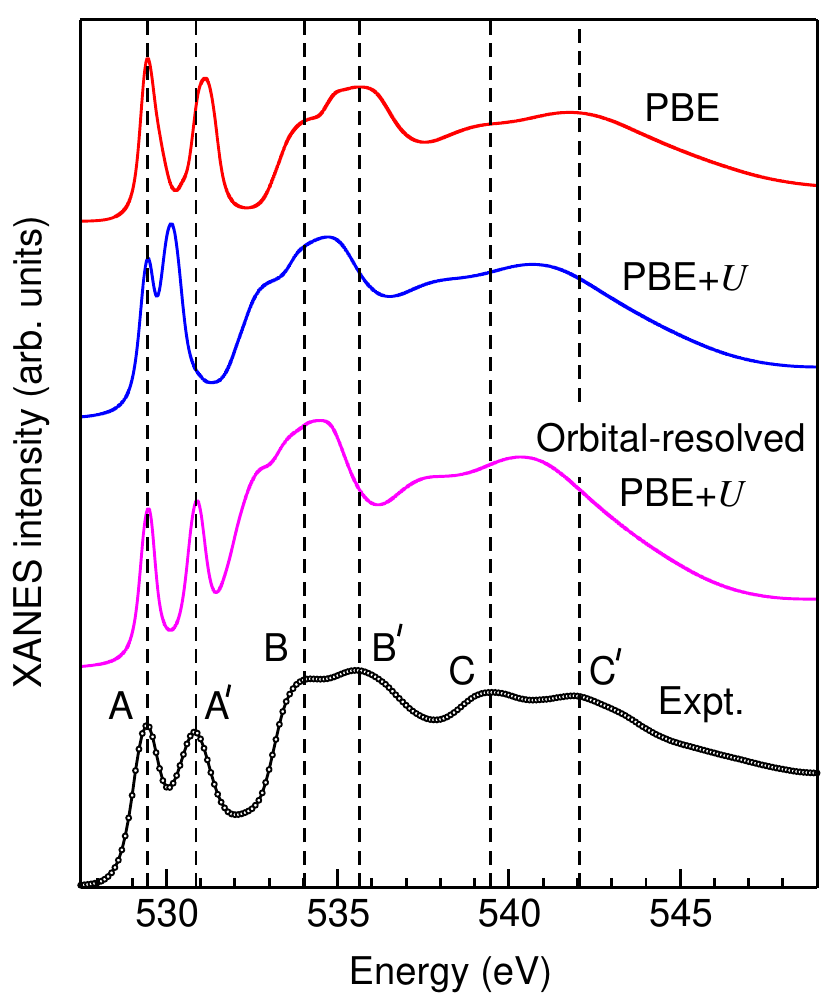}
    \caption{Experimental and calculated O $K$-edge XANES spectra of DyFeO$_3$. 
    The calculated spectra are obtained on top of the PBE, PBE+$U$ (see Table~\ref{tab:Hub_param}), and OR-PBE+$U$ (see Table~\ref{tab:Hub_param_OR}) ground states. The main experimental features are labeled A, A$'$, B, B$'$, C, and C$'$.}
    \label{fig:XANES}
\end{figure}

We now compare the calculated O $K$-edge XANES spectra obtained on top of the PBE, PBE+$U$, and OR-PBE+$U$ ground states with experiment (Fig.~\ref{fig:XANES}). At the O $K$ edge, dipole selection rules allow transitions from the O-$1s$ core level to unoccupied states with O-$2p$ character. The XANES spectrum therefore provides a probe of the O-$2p$-projected unoccupied electronic structure, with the spectral intensity additionally determined by the transition matrix elements. Since O-$2p$ states hybridize with the Fe-$3d$, Dy-$4f$, Dy-$5d$, and higher-energy states, the spectrum provides an indirect probe of the corresponding electronic states and their relative energy scales. We neglect core-hole and electron-hole interaction effects in the present calculations.

The experimental spectrum exhibits six main features, labeled A, A$'$, B, B$'$, C, and C$'$. Following Ref.~\cite{Chiang:2011}, A and A$'$ have been attributed primarily to O-$2p$ and Fe-$3d$ hybridization, B and B$'$ to O-$2p$ and Dy-$5d$ hybridization, and C and C$'$ to hybridization with higher-energy Fe-$4s$, Fe-$4p$, and Dy-$6s$ states. We now assess these assignments and the corresponding theoretical description.

The A and A$'$ features are reproduced by all three calculations, but their separation differs substantially. In PBE, the splitting is somewhat overestimated, consistent with the excessively large Fe-$3d$ crystal-field splitting of approximately $1.7$\,eV discussed in Sec.~\ref{sec:PBE_andPBE+U}. In PBE+$U$, the A--A$'$ separation becomes too small because the shell-averaged Hubbard correction reduces the Fe-$3d$ crystal-field splitting to approximately $0.7$\,eV. In contrast, OR-PBE+$U$ reproduces the separation of A and A$'$ very well. This agreement reflects the orbital-resolved treatment of the Fe-$3d$ states, which was calibrated against HSE06 and yields a crystal-field splitting close to the experimental value. The corresponding O-$2p$ spectral weight arises predominantly from hybridization with Fe-$3d$ states, while the contribution of the Dy-$4f$ states in this energy range is absent.

The B and B$'$ features are predominantly associated with O-$2p$ and Dy-$5d$ hybridization. In PBE+$U$ and OR-PBE+$U$, the Dy-$4f$ states also lie in the same energy range, particularly in the latter where the Dy-$4f$ and Dy-$5d$ manifolds substantially overlap. Nevertheless, no pronounced additional Dy-$4f$ fingerprint is apparent in the O $K$-edge spectrum. This can be attributed to the localized nature of the Dy-$4f$ orbitals, which limits their spatial overlap and hybridization with O-$2p$ states compared with the more extended Dy-$5d$ orbitals. Thus, the B and B$'$ features remain predominantly Dy-$5d$ derived, while any Dy-$4f$ contribution is likely weaker and masked by the broader spectral background. Interestingly, the positions of B and B$'$ are better reproduced by PBE than by PBE+$U$ or OR-PBE+$U$. This discrepancy likely reflects the absence of an explicit Hubbard correction for the Dy-$5d$ states. An additional correction to this manifold could potentially improve the agreement with experiment by shifting the Dy-$5d$-derived spectral weight to higher energies. Such a correction is, however, not currently possible within our OR-PBE+$U$ implementation because simultaneous orbital-resolved corrections of the Dy-$5d$ and Dy-$4f$ channels are not supported.

The higher-energy C and C$'$ features are associated with hybridization involving more delocalized Fe-$4s$, Fe-$4p$, and Dy-$6s$ states~\cite{Chiang:2011}. Their positions are again better reproduced by PBE, than by PBE+$U$ or OR-PBE+$U$. This behavior results from the selective nature of the Hubbard correction: the Fe-$3d$ and Dy-$4f$ states are explicitly modified, while the higher-energy states are affected only indirectly through self-consistency. A direct XANES calculation based on HSE would provide a useful reference for this energy range (since HSE directly acts on all states), but is presently not available within our computational framework.

Overall, the XANES comparison demonstrates that OR-PBE+$U$ provides the most accurate description of the low-energy A and A$'$ features and, in particular, of the Fe-$3d$ crystal-field splitting. Standard PBE+$U$, despite improving the fundamental band gap, gives a substantially less accurate description of these spectroscopic features. At higher energies, PBE yields better peak positions, while both Hubbard-corrected approaches exhibit a contraction of the spectrum, indicating that the Dy-$5d$ and higher-energy states may require additional corrections. Finally, despite their pronounced influence on the electronic structure and magnetic properties of DyFeO$_3$~\cite{Stroppa:2010,Ameri:2021,Cui:2024}, the Dy-$4f$ states leave no clear fingerprint in the O $K$-edge spectrum, consistent with their strongly localized character and consequently weaker hybridization with O-$2p$ states.

\section{Conclusions}
\label{sec:conclusions}

We have performed a first-principles investigation of the electronic structure of DyFeO$_3$ using PBE, PBE+$U$, HSE, and OR-PBE+$U$, followed by calculations of the O $K$-edge XANES spectrum. The Hubbard parameters of PBE+$U$ were determined from first principles using DFPT, while the orbital-resolved Hubbard parameters were tuned to reproduce the HSE06 electronic structure. HSE06 was adopted as the reference after assessing the dependence of the electronic structure on the exact-exchange fraction and comparing the results with available experimental data.

Our results show that standard PBE+$U$, despite substantially improving the fundamental band gap, does not provide a uniformly accurate description of the electronic states relevant to the O $K$-edge XANES spectrum. In particular, the shell-averaged Hubbard correction leads to an excessively small Fe-$3d$ crystal-field splitting of 0.7\,eV. HSE06 provides a more balanced description of the relevant electronic energy scales, yielding a fundamental gap of $\sim2.3$\,eV and an Fe-$3d$ crystal-field splitting of $\sim1.4$\,eV, both in good agreement with the available experimental data. We therefore use the HSE06 electronic structure as a reference for parametrizing OR-PBE+$U$.

We find that resolving the Hubbard parameters at the level of individual orbital submanifolds is essential for reproducing the HSE06 electronic structure. The resulting OR-PBE+$U$ calculation captures the main features of the HSE06 PDOS and, importantly, reproduces the experimentally observed separation of the two lowest-energy O $K$-edge features associated with O-$2p$ and Fe-$3d$ hybridization. This demonstrates that orbital resolution provides the additional flexibility required to describe the localization and energies of the Fe-$3d$ and Dy-$4f$ states while retaining the computational efficiency of a Hubbard-corrected functional.

The comparison with experiment further shows that the O $K$-edge XANES spectrum is particularly sensitive to the Fe-$3d$ crystal-field splitting, whereas the localized Dy-$4f$ states leave no distinct spectral fingerprint despite their strong influence on the electronic and magnetic properties of DyFeO$_3$. At higher energies, discrepancies between theory and experiment point to the need for a more accurate treatment of the Dy-$5d$ and other delocalized unoccupied states, which are affected by exact exchange in HSE but are not explicitly corrected within the present OR-PBE+$U$ framework.

Overall, this work demonstrates a practical strategy for obtaining XANES spectra when hybrid-functional calculations are not directly available: an advanced electronic-structure method can first be used to establish a physically motivated reference, which is then mapped onto an orbital-resolved DFT+$U$ description compatible with the XANES formalism. The results further highlight that reproducing the band gap alone is insufficient for reliable XANES spectroscopic predictions; the orbital character and relative energy separation of the relevant unoccupied states must also be described accurately. At the same time, the present results underscore the need for further developments of first-principles methods and computational implementations to enable more quantitative and accurate simulations of XANES spectra in complex materials.

\section*{Acknowledgments}

This research was supported by the NCCR MARVEL, a National Centre of Competence in Research, funded by the Swiss National Science Foundation (Grant No. 205602). I.T. acknowledges support from the Swiss National Science Foundation (Grant No.~200021-227641 and No.~200021-236507). B.B. and C.W.S. acknowledges support by the Swiss National Science Foundation (Project No. 200020-169393). Computer time was provided by the Swiss National Supercomputing Centre (CSCS) under projects No.~lp18, s1326, and s1335. Part of this work was performed at the Surface/Interface Microscopy (SIM) beamline of the Swiss Light Source (SLS), Paul Scherrer Institut (PSI), Villigen, Switzerland. 

\section*{Appendix: Orbital-resolved Hubbard parameters from first principles}

Here we present our results for determining orbital-resolved Hubbard $U$ parameters from first principles using linear-response theory based on finite differences, following the scheme described in Ref.~\cite{Macke:2024}. We use the same computational setup as in Sec.~\ref{sec:comput_details} and apply small perturbations of strength $\alpha = \pm 0.05$ and $\pm 0.10$\,eV to the PBE ground state of DyFeO$_3$, targeting the spin-orbital groups defined in Table~\ref{tab:Hub_param_OR}. The calculations are performed in a ``one-shot'' fashion, i.e., without the iterative self-consistent response procedure employed in Ref.~\cite{Timrov:2021}. We record the resulting changes in the occupations of the targeted groups both after the first self-consistent-field iteration and after convergence, yielding the noninteracting and interacting response matrices, $\chi_0$ and $\chi$, respectively. The orbital-resolved Hubbard $U$ parameters are then obtained from Eq.~\eqref{eq:Ucalc}, generalized to the orbital-resolved case as described in Ref.~\cite{Macke:2024}. Since our aim is to assess whether the first-principles approach yields qualitatively different parameters from those obtained by fitting to HSE06, we do not employ larger supercells and perform all calculations using the cell shown in Fig.~\ref{fig:Crystal_structure}(b).

\begin{table}[b!]
    \renewcommand{\arraystretch}{1.3}
    \centering
    \caption{Orbital-resolved Hubbard $U$ parameters obtained from linear-response calculations~\cite{Macke:2024}. The values shown correspond to one representative Fe ion and one representative Dy ion. The other Fe and Dy ions belonging to the oppositely polarized magnetic sublattice are symmetry equivalent to the respective ions shown here, with the $U$ values for the spin-up and spin-down channels interchanged.}
    \begin{tabular}{l|c|c|c|c|c|c}
    \hline\hline
     \multirow{2}{*}{Manifold} & \multicolumn{3}{c|}{Fe-$3d$} & \multicolumn{3}{c}{Dy-$4f$} \\ \cline{2-7}
        & $3d^\uparrow_{(1\text{--}5)}$ & $3d^\downarrow_{(1\text{--}3)}$ & $3d^\downarrow_{(4,5)}$ & $4f^\uparrow_{(6,7)}$ & $4f^\uparrow_{(1\text{--}5)}$ & $4f^\downarrow_{(1\text{--}7)}$ \\ \hline
     $U$ (eV) & - & $6.0$ & $11.4$ & $7.8$ & $13.2$ & - \\
    \hline\hline
    \end{tabular}
    \label{tab:OR-U-LRcDFT}
\end{table}

The resulting orbital-resolved Hubbard parameters are listed in Table~\ref{tab:OR-U-LRcDFT}. For the Fe-$3d$ states, the calculated $U=11.4$\,eV for the $3d^\downarrow_{(4,5)}$ ($e_g$) manifold is in very good agreement with the fitted value of $11.2$\,eV in Table~\ref{tab:Hub_param_OR}. The calculated value for the $3d^\downarrow_{(1\text{--}3)}$ ($t_{2g}$) manifold, $U=6.0$\,eV, is lower than the fitted value of $7.6$\,eV, but remains of the same order of magnitude. This difference is nevertheless important because the crystal-field splitting between the $t_{2g}$ and $e_g$ states is sensitive to the corresponding Hubbard parameters. As discussed below, using the first-principles values results in an overestimation of this splitting. We do not compute $U$ for the $3d^\uparrow_{(1\text{--}5)}$ manifold because these states are fully occupied, for which the present formulation of the linear-response approach is not applicable~\cite{Yu:2014}.

For the Dy-$4f$ states, the calculated $U=13.2$\,eV for the unoccupied $4f^\uparrow_{(1\text{--}5)}$ manifold is qualitatively consistent with the large fitted value of $9.8$\,eV, although it is more than $3$\,eV larger. In contrast, the occupied $4f^\uparrow_{(6,7)}$ manifold yields a finite value of $U=7.8$\,eV, whereas $U=0$ was required for these states in the HSE06-based calibration. As discussed in the main text, applying such a large Hubbard correction to the $4f^\uparrow_{(6,7)}$ states substantially shifts their spectral weight away from the VBM and therefore changes the character of the VBM. As for the fully occupied Fe-$3d$ states, we do not compute $U$ for the fully occupied $4f^\downarrow_{(1\text{--}7)}$ manifold because the present linear-response formulation is not applicable in such cases~\cite{Yu:2014}.

\begin{figure}[t]
    \centering
    \includegraphics[width=0.98\linewidth]{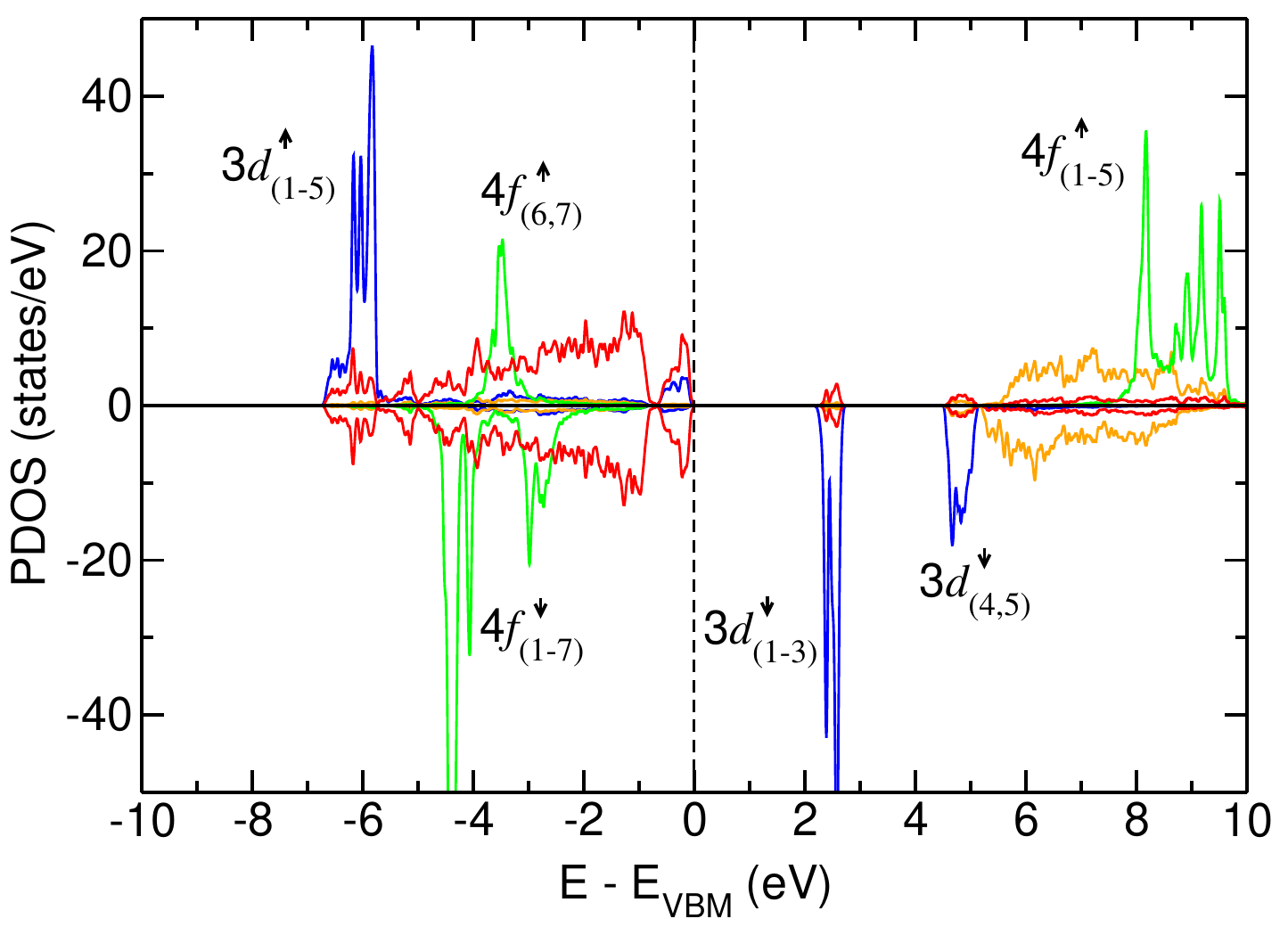}
    \caption{PDOS of DyFeO$_3$ calculated using OR-PBE+$U$ with orbital-resolved Hubbard $U$ parameters listed in Table~\ref{tab:OR-U-LRcDFT}. The electronic structure is calculated for the collinear antiferromagnetic configuration shown in Fig.~\ref{fig:Crystal_structure}(b). The upper and lower halves of each panel show the spin-up and spin-down components, respectively, with the latter plotted with negative sign. For each chemical species, only one of the two symmetry-equivalent magnetic sublattices is shown; the corresponding states of the oppositely polarized sublattice are omitted for clarity. The energy zero is set to the VBM.}
    \label{fig:PDOS_OR-PBE+U_from_LRT}
\end{figure}

The corresponding OR-PBE+$U$ PDOS is shown in Fig.~\ref{fig:PDOS_OR-PBE+U_from_LRT}. It differs substantially from the PDOS obtained using the HSE06-calibrated parameters listed in Table~\ref{tab:Hub_param_OR}, particularly in the distribution of states in the valence region and in the character of the VBM. These differences can be traced primarily to the finite value of $U$ obtained for the Dy-$4f^\uparrow_{(6,7)}$ manifold, which shifts these states away from the VBM. More importantly, the unoccupied Fe-$3d$ states exhibit a substantially larger crystal-field splitting between the $t_{2g}$ and $e_g$ manifolds. The calculated splitting is approximately $2.3$\,eV, compared with the experimental value of approximately $1.4$\,eV. The calculated band gap is approximately $2.2$\,eV, which is close to the experimental value; however, this agreement in the band gap does not imply a consistent description of the electronic structure because the VBM has a different orbital character. The relatively large $U$ value obtained for the unoccupied Dy-$4f^\uparrow_{(1\text{--}5)}$ manifold also shifts these states to higher energies compared with the HSE06 reference. This shift is expected to affect the unoccupied-state distribution and, consequently, the spectral features associated with the O $K$ edge, particularly the first two peaks. 

Overall, these results show that the first-principles determination of orbital-resolved Hubbard parameters using the present linear-response formulation provides useful insight into the orbital dependence of the Hubbard correction. However, the resulting parameters do not reproduce the HSE06 electronic structure quantitatively and consistently across the relevant occupied and unoccupied manifolds. In particular, the treatment of the fully occupied manifolds and the resulting values for the Dy-$4f^\uparrow_{(6,7)}$ and empty Fe-$3d$ submanifolds lead to substantial differences in the VBM character and crystal-field splitting. Thus, within the present formulation and computational setup, the first-principles orbital-resolved $U$ parameters do not provide an equally accurate description of the electronic structure as the parameters calibrated against the HSE06 reference.


\begin{thebibliography}{89}
\bibitem [{Ber()}]{Bertaut_explanation}%
  \BibitemOpen
  \href@noop {} {}\bibinfo {note} {{Bertaut's notation describes the symmetry
  of the magnetic ordering separately for each Cartesian component of the
  magnetic moments. For example, in the $G_xA_yF_z$ configuration of the Fe
  sublattice, the $x$ component follows a $G$-type antiferromagnetic (AFM)
  arrangement, in which nearest-neighbor Fe moments are antiparallel. The $y$
  component follows an $A$-type AFM arrangement, characterized by ferromagnetic
  alignment within the relevant crystallographic planes and antiferromagnetic
  alignment between adjacent planes. The $z$ component follows an $F$-type
  arrangement, corresponding to a ferromagnetic alignment of the Fe moments and
  therefore giving rise to a net ferromagnetic component along the $z$
  direction.}}\BibitemShut {Stop}%
\end{thebibliography}

%

\end{document}